%% file: proc_main.tex
\documentclass[prc,showpacs,floatfix]{revtex4}
\usepackage{bm}
\usepackage{dcolumn}
\usepackage{graphicx}
\usepackage{amsmath}
\usepackage{bbold}
\newcommand{\beq}{\begin{equation}}
\newcommand{\eeq}{\end{equation}}
\newcommand{\beqa}{\begin{eqnarray}}
\newcommand{\eeqa}{\end{eqnarray}}
\newcommand{\beqas}{\begin{eqnarray*}}
\newcommand{\eeqas}{\end{eqnarray*}}
\begin{document}
\title{
\hfill{\small {\bf MKPH-T-08-09}}\\
{\bf Spin degrees and Polarization Observables in Electromagnetic Reactions}
\footnote{Lectures delivered at the DAE-BRNS Workshop on Hadron Physics, AMU,
Aligarh, India, Feb.\ 18-23, 2008.}}
\author{Hartmuth Arenh\"ovel}
\affiliation{
Institut f\"ur Kernphysik,
Johannes Gutenberg-Universit\"at Mainz, D-55099 Mainz, Germany}
\date{\today}
\begin{abstract}
After a brief introduction into the general importance of polarization
observables for the analysis of a reaction, the basic density matrix formalism
for the description of polarization phenomena is outlined and
illustrated by explicit examples for spin one-half and one. Then various
electromagnetic processes like photoreactions and electron scattering are
discussed in some detail. Finally, the question of how to determine a complete
set of polarization observables providing the maximum information on a
reaction is considered. 
\end{abstract}
\maketitle



\input{proc_introduction}
\input{proc_density_matrix}

\input{proc_em_reactions}

\input{proc_complete_sets}

\input{proc_conclusions}

\input{proc_ref}
\end{document}

%% file: proc_introduction.tex
\section{Introduction and Motivation }

Historically the notion of ``polarization'' has had various although related
meanings~\cite{Rob74}: 
\begin{description}
\item{(i)}
Quite early the term ``polarity'' for the description of the two sides of a
magnet was used in order to characterize the separation of magnetic poles. It
thus denotes a directedness associated with respect to the orientation of a
magnetic dipole (compass). 
\item{(ii)}
Probably this term led {\sc E.L.\ Malus} (1811) to introduce the term
``polarization'' in order to describe the property of light waves of
exhibiting a type of vibrations transversal to the direction of propagation
like the vibrations of a string. It was first discovered already by {\sc Ch.\
  Huygens} in 1690, although in those times the nature of the vibrations was
not clear. Since then this property is called ``transverse polarization'',
expressing a directedness. 
\item{(iii)}
Also the separation of charges in a dielectric medium or the orientation of
randomly oriented permanent electric dipoles by an external
electric field is called polarization, which also means the creation of a
preferred direction in the medium. 
\item{(iv)}
In more recent times with the advent of quantum mechanics the term
``polarization'' is used in a much more general sense in order to describe the
orientation of particles, including massless ones, possessing a nonvanishing
spin. In fact, all the above mentioned phenomena are  related to such spin
degrees of freedom. 
\end{description}
Therefore, in the present lectures I will use ``polarization'' in this general
sense with respect to spin degrees of freedom and associated phenomena. 

It is well known that the spin of a particle is an important dynamical property
which often plays a significant role in the dynamics of a reaction. Moreover,
whether a particle possesses integer or half-integer spin determines its
statistical behavior in an ensemble of identical particles, obeying either
{\sc Bose-Einstein} or {\sc Fermi-Dirac} statistics. The central question
which I would like to discuss is ``How can one exploit such spin degrees of
freedom?''. 
 
Let us consider, for example, a reaction induced by two particles of the type
\beqa
a+b\rightarrow c+d+\dots
\eeqa
Let us further assume that the initial system is prepared in a pure state
$|\psi_{i,M_a,M_b}\rangle$ with definite spin projections $M_a$ and
$M_b$ of the initial particles and, furthermore, that the detector selects
ideally another pure final state $|\psi_{f,M_c,M_d,\dots}\rangle$. Then
the cross section is determined by the absolut square of the corresponding
reaction matrix $\widehat T$, leaving aside kinematical factors,
\beqa
\sigma(a+b\rightarrow c+d+\dots) \propto 
|\langle \psi_{f,M_c,M_d,\dots}|\widehat
 T|\psi_{i,M_a,M_b}\rangle|^2\,. 
\eeqa
Spin degrees could then be investigated by varying the projections
$M_a$ and $M_b$ of the initial and $M_c,M_d,\dots$ of the final
states, and thus study how the cross section changes. This ideal situation,
however, is almost never fulfilled, because (i) the initial preparation
results in general in a statistical ensemble of a subset of different states,
and (ii) similarly the detector is sensitive to another statistical ensemble
of final states. In other words, in general one only has incomplete
information about the initial and final states of the system. 

In the simplest case, both are an incoherent mixture of possible initial or
final states, respectively, with equal probability, and then the cross section
is given by averaging over the initial and summing over the final ensemble,
e.g.\ states with different spin projections, yielding for the cross section
\beqa
\sigma(a+b\rightarrow c+d+\dots) \propto 
\sum_{M_a,M_b}\sum_{M_c,M_d,\dots}
|\langle \psi_{f,M_c,M_d,\dots}|\widehat
T|\psi_{i,M_a,M_b}\rangle|^2\,. 
\eeqa
This is, for example, the case in photoreactions with unpolarized photons,
which is another way of saying, that the photon source produces photons which
do not have a specific direction of polarization. In this case, any specific
dependence of the reaction on the photon polarization is lost by the
averaging procedure. 

\begin{figure}[h]
\centerline {\includegraphics[scale=.6]{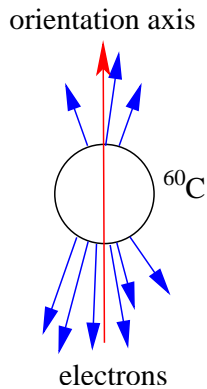}}
\caption{Preferred electron emission opposite to the orientation axis in the
  $\beta$-decay of $^{60}$Co.}
\label{wu_exp}
\end{figure}

In the present lectures I will illustrate how one can employ spin degrees of
freedom for studying details of a reaction. A more detailed account can be
found in a monograph by B.A.\ Robson~\cite{Rob74}. A particularly interesting
feature of exploiting spin d.o.f.\ is the possibility to obtain access to
small but interesting amplitudes which usually are buried under dominant
contributions in an unpolarized experiment. In order to illustrate this
feature, I would like to mention two examples: 
\begin{description}
\item{(i)} 
The first measurement of parity violation in $\beta$-decay of oriented
$^{60}$Co~\cite{WuA57}, where one measures the correlation between the
orientation axis of $^{60}$Co and the direction of the emitted electrons,
which is proportional to the scalar product between the spin of $^{60}$Co and
the electron momentum, a pseudoscalar quantity generated by parity
violation. It turned out that the electrons are predominantly emitted opposite
to the orientation axis (see Fig.~\ref{wu_exp}). 
\item{(ii)} 
The determination of the electric form factor of the neutron in quasielastic
electron scattering with longitudinally polarized electrons and measuring the
polarization of the outgoing neutron $d(\vec{e},e'\vec{n})p$ or using a vector
polarized deuteron target $\vec{d}(\vec{e},e'n)p$. I will discuss this case in
some detail later.
\end{description}

%% file: proc_density_matrix.tex
\section{The Density Matrix }

I will now briefly introduce the concept of the {\it density matrix} as a
convenient tool for the description of statistical ensembles (for reviews
see~\cite{Fan57,THa61}). Let us consider 
a quantum mechanical system which is described by a complete set of states $\{ 
|\alpha\rangle; \alpha\in \{\mbox{index set}\}\}$. Furthermore, an observable
is represented by an hermitian operator $\widehat \Omega$. Ideally, a
measurement of an observable determines matrix elements between pure states 
\beqa
\Omega_{fi}=\langle\psi_f|\widehat\Omega|\psi_i\rangle\,.
\eeqa
However, as already mentioned, this ideal situation is almost never realized
because one deals in general with a large number of identical systems with
incomplete information on which states they occupy, so-called {\it mixed
  states}. Therefore, statistical methods are needed. 
This leads to the introduction of the {\it density matrix} as the appropriate
tool for representing pure and mixed states on an equal footing. 

For illustration I will consider the expectation value of an observable
$\widehat\Omega$ for both cases, pure and mixed states. For a pure state,
represented by $|\psi\rangle$ the expectation value simply is given by
\beqa
\langle\Omega\rangle_{|\psi\rangle}=\langle\psi|\widehat\Omega|\psi\rangle\,,
\eeqa
whereas for a mixed state, represented by an incoherent superposition of a set
of normalized states $\{|\psi_1\rangle,...,|\psi_n\rangle\}$ with probabilities
$\{p_1,...,p_n\}$, where $\sum_i p_i=1$, one has 
\beqa
\langle\Omega\rangle_{\{|\psi_1\rangle,...,|\psi_n\rangle\}}=
\sum_{i=1}^np_i\langle\psi_i|\widehat\Omega|\psi_i\rangle\,.
\eeqa
To give an example, let me consider the case of unpolarized electrons
represented by the two possible spin projections on a given quantization axis 
$\{|\psi_{1/2}\rangle=|\pm\frac{1}{2}\rangle\}$ with equal probabilities 
$p_1=p_2=\frac{1}{2}$. Then the expectation value is given by  
\beqa
\langle\widehat\Omega\rangle_{\{|1/2\rangle,|-1/2\rangle\}}=
\sum_{m=-\frac{1}{2},\frac{1}{2}}\frac{1}{2}
\langle m|\widehat\Omega|m\rangle\,.
\eeqa
Instead of always distinguishing between these two types of states, the
density matrix formalism offers a unified description for both cases. In order
to introduce the density matrix, I will again consider 
the two cases of pure and mixed states. For a pure state $|\psi\rangle$, which
can be expanded in terms of a complete set of basis states
\beqa
|\psi\rangle &=& \sum_\alpha |\alpha\rangle\langle\alpha|\psi\rangle\,,
\eeqa
the expectation value of $\widehat\Omega$ is given by
\beqa
{ \langle\widehat\Omega\rangle_{|\psi\rangle}}&=&\sum_{\alpha\alpha'}
\langle\psi|\alpha\rangle\langle\alpha|\Omega|\alpha'\rangle
\langle\alpha'|\psi\rangle\nonumber\\
&=&\sum_{\alpha\alpha'}\langle\alpha|\widehat\Omega|\alpha'\rangle
\langle\alpha'|\psi\rangle\langle\psi|\alpha\rangle\,.
\eeqa
Introducing now the {\it density matrix operator }
\beq
\widehat\rho=|\psi\rangle\langle\psi|\,,
\eeq
one obtains
\beqa
\langle\widehat\Omega\rangle_{|\psi\rangle}&=&
\sum_{\alpha}\langle\alpha|\widehat\Omega{\widehat\rho}|\alpha\rangle
={ \mbox{Tr}(\widehat \Omega\widehat\rho)}\,.
\eeqa
Turning to a mixed state described by the set
$\{|\psi_1\rangle,...,|\psi_n\rangle\}$ with probablities $\{p_1,\dots,p_n\}$
and expanding each member of the set according to
\beq
|\psi_i\rangle = \sum_\alpha |\alpha\rangle\langle\alpha|\psi_i\rangle\,,
\eeq
one finds now for the expectation value
\beqa
{ \langle\widehat\Omega\rangle_{\{|\psi_1\rangle,...,|\psi_n\rangle\}}}&=&
\sum_{i}p_i\langle\psi_i|\widehat\Omega|\psi_i\rangle\nonumber\\
&=&\sum_{i}p_i\sum_{\alpha\alpha'}
\langle\psi_i|\alpha\rangle\langle\alpha|\Omega|\alpha'\rangle
\langle\alpha'|\psi_i\rangle\nonumber\\
&=&\sum_{\alpha\alpha'}\langle\alpha|\widehat\Omega|\alpha'\rangle
\langle\alpha'|\sum_{i}|\psi_i\rangle p_i\langle\psi_i|\alpha\rangle\nonumber\\
&=&{ \mbox{Tr}(\widehat \Omega\widehat\rho)}\,.
\eeqa
In the last step one has introduced as {\it density matrix operator}
\fboxsep2mm
\beqa
\fbox{$\displaystyle \widehat\rho=\sum_{i}|\psi_i\rangle p_i\langle\psi_i|.$}
\eeqa
Obviously, the case of a pure state is contained in this definition for
$p_i=\delta_{in_0}$ with $n_0\in\{1,...,n\}$.  

\subsection{Properties of the Density Matrix }
Here I will list the most important properties of the density matrix operator.
\begin{description}
\item{(i)}
\framebox{$\widehat\rho^\dagger=\widehat\rho$}; i.e.\ $\widehat\rho$ is
hermitean, because $|\psi\rangle^\dagger=\langle\psi|$ and thus 
$(|\psi\rangle\langle\psi|)^\dagger=\langle\psi|^\dagger|\psi\rangle^\dagger
=|\psi\rangle\langle\psi|$.
\item{(ii)}
\framebox{$\mbox{Tr}\,\widehat\rho=1$}; because in terms of a complete set one
has 
\beqa
\mbox{Tr}\,\widehat\rho&=&
\sum_{\alpha}\langle\alpha|\widehat\rho|\alpha\rangle\nonumber\\
&=&\sum_{\alpha}\sum_{i}\langle\alpha|\psi_i\rangle
p_i\langle\psi_i|\alpha\rangle
=\sum_{i}p_i\sum_{\alpha}\langle\psi_i|\alpha\rangle\langle\alpha|\psi_i\rangle
\nonumber\\ 
&=&\sum_{i}p_i\langle\psi_i|\psi_i\rangle=\sum_{i}p_i=1\,.
\eeqa
\item{(iii)}
\framebox{$\rho_{\alpha\alpha}\ge 0$}; i.e.\ all diagonal elements are real and
non-negative, because 
\beqa
\rho_{\alpha\alpha}=\sum_{i}\langle\alpha|\psi_i\rangle
p_i\langle\psi_i|\alpha\rangle=
\sum_{i}p_i|\langle\alpha|\psi_i\rangle|^2\ge 0\,.
\eeqa
Obviously, $\rho_{\alpha\alpha}$ is the probability for finding the state
$|\alpha\rangle$ in the ensemble.
Furthermore, according to (ii) one has for the diagonal elements as upper and
lower limits $1\ge \rho_{\alpha\alpha}\ge 0$. 
\item{(iv)}
\framebox{ $\mbox{Tr}\,\widehat\rho^{\,2}\leq 1 $}; because one finds
\beqa
\mbox{Tr}\,\widehat\rho^{\,2}&=&\sum_{\alpha}\langle\alpha|
\Big(\sum_{i}|\psi_i\rangle p_i\langle\psi_i|\Big)
\Big(\sum_{j}|\psi_j\rangle p_j\langle\psi_j|\Big)|\alpha\rangle\nonumber\\
&=&\sum_{i,j}p_ip_j\langle\psi_i|\psi_j\rangle
\sum_{\alpha}\langle\psi_j|\alpha\rangle\langle\alpha|\psi_i\rangle\nonumber\\
&=&\sum_{i,j}p_ip_j|\langle\psi_i|\psi_j\rangle|^2\,.
\eeqa
Using now Schwartz' inequality
\beqa
|\langle\psi_i|\psi_j\rangle|^2\leq 
\langle\psi_i|\psi_i\rangle\langle\psi_j|\psi_j\rangle=1\,,
\eeqa
one finds finally
\beqa
\mbox{Tr}\,\widehat\rho^{\,2}&\leq&
\Big(\sum_ip_i\Big)^2=(\mbox{Tr}\,\widehat\rho\,)^2=1\,.
\eeqa
Equality is fulfilled if and only if the state is pure, because for a pure
state one has $\widehat \rho =|\psi\rangle\langle\psi|$ where $|\psi\rangle$
is normalized to one, and thus $(\widehat \rho)^2=\widehat \rho$. The inverse
follows from the next property (v).
\item{(v)}
\framebox{$\widehat \rho^{\,2}=\widehat \rho$} if and only if the state is
pure. We only have to show that, if this relation holds, then the density
matrix has the form $\widehat \rho =|\psi\rangle\langle\psi|$. To this end we
note that since $\widehat \rho$ is hermitean one can diagonalize the density
matrix. This means one can find a complete set of states
$\{|\phi_1\rangle,\dots,|\phi_n\rangle,\dots\}$ with $
\langle\phi_i|\widehat \rho\,|\phi_j\rangle=\delta_{ij}\rho_{ii}$. From this
feature follows, using the above relation,
\beqa
\langle\phi_i|\widehat \rho^{\,2}|\phi_j\rangle=\delta_{ij}\rho_{ii}^2
=\langle\phi_i|\widehat \rho|\phi_j\rangle=\delta_{ij}\rho_{ii}\,.
\eeqa
Thus for all $i$ one has the equation $\rho_{ii}^2=\rho_{ii}$ which has the
only solutions $\rho_{ii}=0$ or $\rho_{ii}=1$. In view of the properties
(ii) and (iii) only for one index $n_0$ the solution can be non-vanishing,
i.e.\ $\rho_{ii}=\delta_{in_0}$. This means that $\widehat
\rho=|\phi_{n_0}\rangle\langle\phi_{n_0}|$, q.e.d. 
In this case $\widehat \rho$ is {\it idempotent} or a projection operator onto
the pure state $|\phi_{n_0}\rangle$.  
\item{(vi)}
\framebox{$\langle\widehat\Omega\rangle_\rho=
\mbox{Tr}(\widehat\Omega\widehat\rho)$}; i.e.\ the expectation value of any
operator $\widehat\Omega$ is given by the trace of
$\widehat\Omega\widehat\rho$. This property may also serve as a
definition for the density matrix in an operational approach~\cite{THa61}. 
\item{(vii)}
For a $N$-dimensional density matrix ($N$ basis states) the matrix is
characterized in general by $N^2-1$ real parameters. Only for a pure state
this number reduces to $2N-2$, because then the pure state is a linear
combination of the $N$ basis states with $N$ complex coefficients for which an
overall phase remains arbitrary. 
\item{(viii)}
The density matrix fulfills the equation of motion
\beqa
\fbox{$\displaystyle
\frac{\partial\widehat\rho}{\partial
t}=\frac{i}{\hbar}\,\Big[\widehat\rho,\widehat H\Big],$} 
\eeqa
where $\widehat H$ denotes the hamiltonian of the system.
\end{description}
As next I will discuss briefly the density matrices for the description of
spin degrees of freedom.

\subsection{Density Matrix for Spin 1/2}
For a spin-one-half particle, one has only two basis states, namely
$\{|\frac{1}{2}\, m_s\rangle; m_s=\pm\frac{1}{2}\}$. Therefore, the
corresponding density matrix $\rho^{(1/2)}$ is a $2\times 2$-matrix which can
be represented as a linear combination of the $2\times 2$ unit matrix
$\mathbb{1}_2$ and the {\sc Pauli} spin matrices $\sigma_i$
\beqa
\rho=\frac{1}{2}(\mathbb{1}_2+a_1\sigma_1 +a_2\sigma_2
+a_3\sigma_3)\,,
\eeqa
where the coefficients $a_i$ are real, because both $\rho$ and $\sigma_i$ are
hermitean. For the interpretation of these coefficients one considers the
probability for finding a spin component in the direction of the $i$-axis
\beqa
P_i=\mbox{Tr}(\rho^{(1/2)}\sigma_i)=\frac{1}{2}\sum_j a_j
{\mbox{Tr}(\sigma_i\sigma_j)}&=&a_i\,,
\eeqa
because $\mbox{Tr}(\sigma_i\sigma_j)=2\delta_{ij}$. Therefore, the density
matrix of a ($s=1/2$)-particle has the form\\
\beqa
\fbox{$\displaystyle
\rho^{(1/2)}=\frac{1}{2}(\mathbb{1}_2+\mathbf{P}\cdot\mathbf{\sigma}),$}
\eeqa
where $P_i$ is the probability for finding a particle with spin component
along the $i$-axis in the state described by $\rho^{(1/2)}$.
For a pure state one has $|\mathbf{P}|=1$, while for a partially polarized
particle $|\mathbf{P}|<1$. 

\subsection{Density Matrix for a Photon}
Since a photon has only two independent transverse polarizations, one has
again only two basis states for which two types of representations are used:
\begin{description}
\item{(i)} 
States of linear polarization (cartesian basis)
$\{|\mathbf{e}_x\rangle,|\mathbf{e}_y\rangle\}$ with respect to a coordinate
system with $z$-axis parallel to the photon momentum. 
\item{(ii)} 
States of circular polarization (spherical basis)
$\{|\mathbf{e}_{-1\rangle},|\mathbf{e}_1\rangle\}$, which are related to the
cartesian basis by 
\beqa
|\mathbf{e}_{\pm 1}\rangle=\mp\frac{1}{\sqrt{2}}
(|\mathbf{e}_x\rangle\pm i\,|\mathbf{e}_y\rangle).
\eeqa
\end{description}
Since there are again only two basis states, formally the density matrix
$\rho^{(\gamma)}$ has the same structure as for a spin one-half particle 
\beqa
\rho^{(\gamma)}=\frac{1}{2}(\mathbb{1}_2+\mathbf{P}\cdot \mathbf{\sigma})\,.
\eeqa
However, the meaning of $\mathbf{P}$ is different and depends on the chosen
representation, either cartesian or spherical.

\begin{figure}[h]
\centerline{\includegraphics[scale=.7]{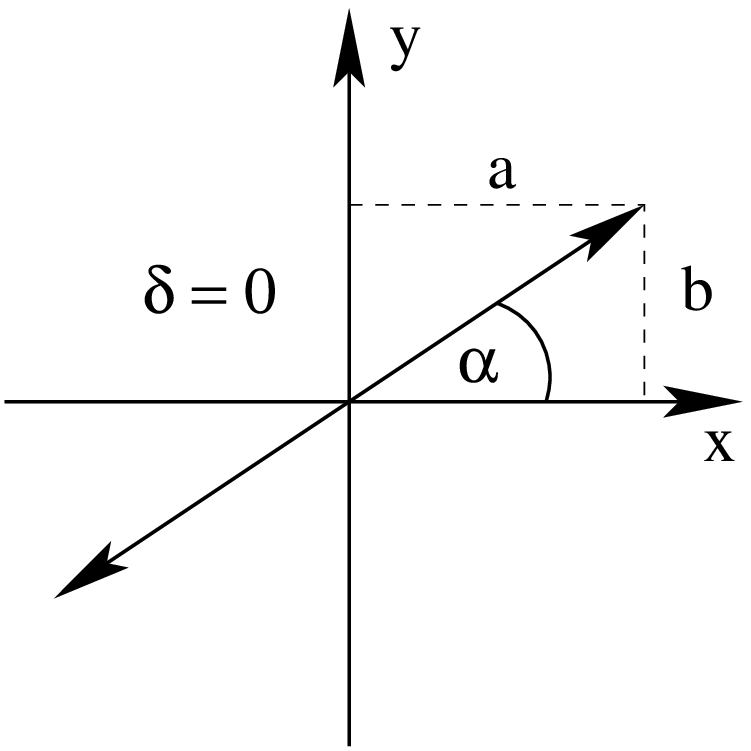}\hspace*{1cm}\includegraphics[scale=.8]{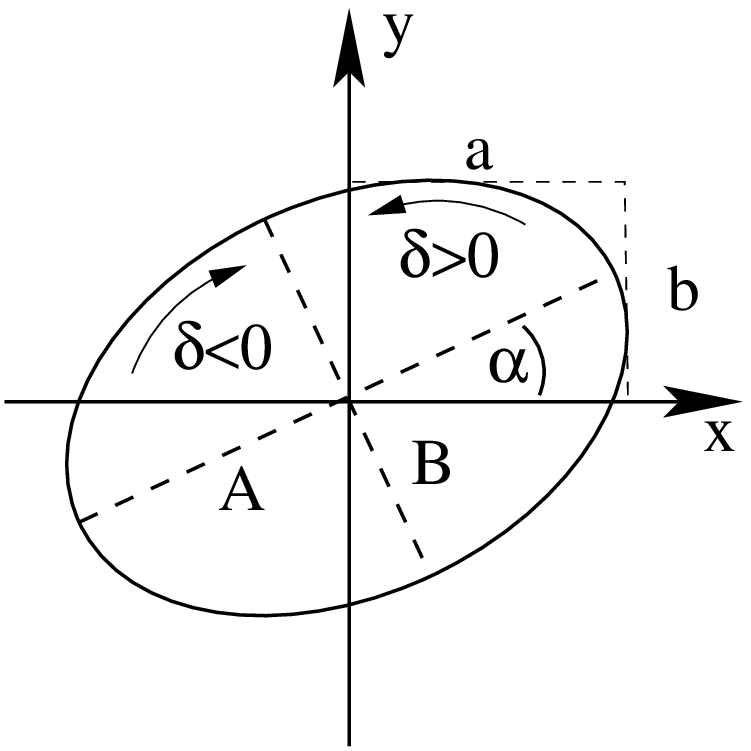}} 
\caption{Left panel: linear polarization; Right panel: elliptic polarization
  of a classical electromagnetic plane wave.} 
\label{stokes}
\end{figure}

At this point I would like to make a digression to classical radiation
physics, where most of the concepts for the description of photon polarization
have been developed. To this end I will consider an electromagnetic plane wave
travelling along the $z$-axis. Its electric field has the form
\beq
\mathbf{E}=\Re e(\mathbf{A}e^{i(kz-\omega t)})\,,
\eeq
with a complex transverse amplitude in the $x$-$y$-plane
$\mathbf{A}=a\mathbf{e}_x +b\,e^{i\delta}\mathbf{e}_y$. For a convenient
description of all possible polarizations {\sc G.G.\ Stokes} has introduced a
set of four parameters, called {\it Stokes parameters}, $\{I,P_1,P_2,P_3\}$
which are defined by 
\beqa
I=a^2+b^2\,,\quad P_1=a^2-b^2\,,\quad P_2=2ab\cos\delta\,,
\quad P_3=2ab\sin\delta\,,
\eeqa
of which only three are independent, because of the condition
\beqa
I^2=P_1^2+P_2^2+P_3^2\,.
\eeqa
Here the intensity is denoted by $I$ and the three parameters $P_i$
characterize the type of polarization, namely linear, circular or elliptic. In
detail one has 
\begin{description}
\item{(i)}
linear polarization for $\delta=0$, i.e.\ $P_3=0$. If $P_2=0$ one has linear
polarization along the $x$-axis if $P_1>0$ or along the $y$-axis if $P_1<0$.
For all other cases, i.e.\  $P_2\neq 0$, one has linear polarization at an
angle $\tan\alpha=b/a$ in the $x$-$y$-plane (see left panel of Fig.~2).
\item{(ii)}
For the case ${ \delta\neq 0}$ one has in general right ($\delta >0$) or left
($\delta <0$) elliptic polarization with the large axis at an angle $\tan
2\alpha=2ab\cos\delta/(a^2-b^2)$ in the $x$-$y$-plane (see right panel of 
Fig.~\ref{stokes}). Circular polarization is a special case if
$\delta=\pm\pi/2$ and $a=b$. In this case the Stokes parameters have the
values $P_1=P_2=0$ and $P_3=\pm I$. 
\end{description}

\begin{figure}
\centerline {\includegraphics[scale=.5]{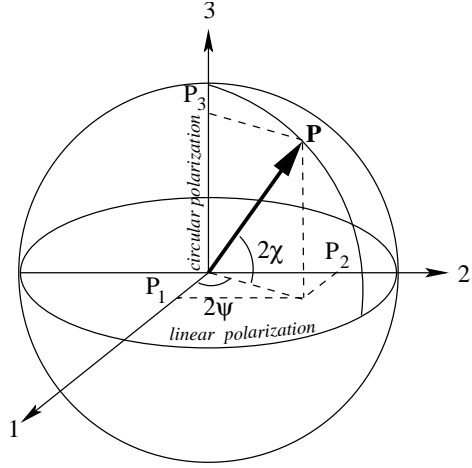}}
\caption{The {\em Poincar\'e Sphere}, a three-dimensional
  representation of the polarization of a classical electromagnetic plane
  wave.}
\label{poincare}
\end{figure}

{\sc H.\ Poincar\'e} has introduced a three-dimenional representation for all
possible polarization states, the so-called {\em Poincar\'e vector}
$\mathbf{P}$, defined by
\beqa
P_1=I\cos 2\chi\cos 2\psi\,,\quad P_2=I\cos 2\chi\sin 2\psi\,,
\quad P_3=I\sin 2\chi\,,
\eeqa
where
\beqa
\tan 2\psi=\frac{2ab\cos\delta}{a^2-b^2}\quad\mbox{and}\quad\tan
2\xi=\tan\delta\sin 2\psi\,.
\eeqa
The Poincar\'e vector has the length $I$ and the spherical angles
$(2\psi,\pi/2-2\xi)$ and thus covers a sphere with radius $I$ (see
Fig.~\ref{poincare}). In other words, each point on the Poincar\'e sphere
characterizes a specific polarization of an electromagnetic plane wave.

Now I will return to the quantum description of photon polarization.
Analogous to classical radiation, the most general photon polarization state
is given by 
\beqa
|\mathbf{e}\rangle=a\,|\mathbf{e}_x\rangle +b\,e^{i\delta}\,
|\mathbf{e}_y\rangle\,,
\quad\mbox{with}\quad  a,b\mbox{ real}\,,\quad\mbox{and}\quad a^2+b^2=1\,,
\eeqa 
and $\delta$ denotes a phase. It describes the same types of polarization, 
linear, circular and elliptic, as in the classical case. The only difference
is that now the Poincar\'e vector has length one. With respect to the
cartesian basis, the corresponding density matrix has the form
\beqa
\rho_{cart}^{(\gamma)}&=&
\left(\begin{matrix}
a^2 & ab e^{-i\delta} \cr
ab e^{i\delta}& b^2\cr
\end{matrix}\right)
=\frac{1}{2}\,\left(\begin{matrix}
1+a^2-b^2 & 2ab(\cos{\delta} -i\sin{\delta})\cr
2ab(\cos{\delta} +i\sin{\delta}) & 1-a^2+b^2\cr
\end{matrix}\right)\nonumber\\
&=&\frac{1}{2}(\mathbb{1}_2+P_2\,\sigma_x +P_3\,\sigma_y
+P_1\,\sigma_z)\,.
\eeqa
Thus one finds with $(P_x,P_y,P_z)=(P_2, P_3, P_1)$ for the photon
polarization density matrix in the cartesian basis
\beqa
\fbox{$\displaystyle
\rho_{cart}^{(\gamma)}=
\frac{1}{2}(\mathbb{1}_2+P_x\sigma_x+P_y\sigma_y+P_z\sigma_z)\,.$}
\eeqa
For a pure polarization state one has $|\mathbf{P}|=1$, whereas for partially
polarized photons $|\mathbf{P}|<1$. 

As next I will consider the transformation to the spherical basis. The
transformation of the density matrix to this basis is given by
\beqa
\rho_{sph}^{(\gamma)}={\cal U}^*\rho_{cart}^{(\gamma)}
\,{\cal U}^T\quad\mbox{with}\quad
{\cal U}=\frac{1}{\sqrt{2}}\left(\begin{array}{rr}
-1 &\quad -i\\1 &\quad -i\\\end{array}\right)\,,
\eeqa
yielding
\beqa
\rho_{sph}^{(\gamma)}&=&
\frac{1}{2}\,\left(\begin{matrix}
1+2ab\sin{\delta} & b^2-a^2+ 2iab\cos{\delta}\cr
b^2-a^2- 2iab\cos{\delta} & 1-2ab\sin{\delta}\cr
\end{matrix}\right)\,.
\eeqa
Thus, it results in the same formal expression as for the cartesian basis but
with a different polarization vector $(P_x,P_y, P_z)=(-P_1,-P_2, P_3)$ and a
different interpretation. Again $P=|\mathbf{P}|$ denotes the fraction
of total polarization, $P_c^\gamma=P_z$ describes right ($P_c^\gamma>0$) or
left ($P_c^\gamma<0$) circular polarization. With repect to the other
components $P_x$ and $P_y$ it is useful to consider a rotation in the
$x$-$y$-plane by $\phi$ which results in a transformation of the density
matrix $\rho_{sph}^{(\gamma)}\to \rho_{sph}^{(\gamma)\prime}$ with 
\beqa
\rho_{sph,\lambda'\lambda}^{(\gamma)\prime}=e^{i(\lambda'-\lambda)\phi}\,
\rho_{sph,\lambda'\lambda}^{(\gamma)}.
\eeqa
Thus, by a proper rotation in the $x$-$y$-plane one can reach $P_y=0$
and then $P_l^\gamma=P_x$ denotes linear polarization along the $x$-axis
($P_l^\gamma<0$) or along the $y$-axis ($P_l^\gamma>0$).

\subsection{Density Matrix for a Massive Spin-1 Particle}
In contrast to the photon, one has three basis states, namely
$\{|1\,m\rangle;\,m=0,\pm 1\}$. Thus the density matrix $\rho_{m'm}^{(1)}$ is
a $3\times3$-matrix in spin-1 space which can be decomposed into 9 independent
$3\times3$-matrices, the unit matrix $\mathbb{1}_3$ and 8 traceless hermitean
matrices. In view of angular momentum algebra it is convenient to chosen them
as irreducible tensors under rotations of rank one and two, i.e.\ three vector
components $\Omega^{[1]}_M$ (transforming like a spin 1) and five tensor
components $\Omega^{[2]}_M$ (transforming like a spin 2), defined by their
reduced matrix elements 
\beqa
\langle 1\|\Omega^{[I]}\| 1\rangle=\sqrt{3}\widehat I\,,
\eeqa
with the notation $\widehat I=\sqrt{2I+1}$.
Accordingly, the density matrix $\rho^{(1)}$ is governed by eight independent
parameters: 3 for vector polarization ($P_{1M}$) and 5 for tensor polarization 
($P_{2M}$)
\beqa
\fbox{$\displaystyle
\widehat\rho^{(1)}=\frac{1}{3}\,\Big(\mathbb{1}_3+\sum_{I=1,2}\sum_M (-)^M 
\Omega^{[I]}_MP_{IM}\Big).$}
\eeqa
Under spatial rotations $P_{1M}$ transforms like a vector and $P_{2M}$ like a
tensor of rank two with respect to a spherical representation. The relation to
the cartesian vector and tensor components are 
\beqa
P_{1M}&=&\sqrt{\frac{3}{2}}\,(\delta_{M0}P_z-\frac{M}{2}(P_x+iM\,P_y))\\
P_{20}&=&\frac{1}{\sqrt{2}}P_{zz}\,,\quad P_{2\pm 1}=\mp\frac{1}{\sqrt{3}}\,
(P_{xz}\pm i\,P_{yz})\,,\quad P_{2\pm
2}=\frac{1}{2\sqrt{3}}\,(P_{xx}-P_{yy})\pm\frac{2i}{\sqrt{3}}P_{xy}\,. 
\eeqa
For a diagonal density matrix $\rho_{m'm}^{(1)}=\delta_{m'm}p_m$, with $p_m$
denoting the probability for finding the particle in a state $|1m\rangle$ with
respect to an orientation axis parallel to the $z$-axis, the orientation
parameters have a particularly simple form
\beqa
P_{1M}=\delta_{M0}P^0_1\,,\quad P_{2M}=\delta_{M0}P^0_2\,,
\eeqa
where
\beqa
P^0_1=\sqrt{\frac{3}{2}}(p_1-p_{-1})\,,\quad 
P^0_2=\sqrt{\frac{1}{2}}(p_1+p_{-1}-2p_0)
=\sqrt{\frac{1}{2}}(1-3p_0)\,.
\eeqa
For an arbitrary direction of the orientation axis the density matrix can be
obtained from the foregoing by a simple rotation. Then the spin-1 density
matrix is characterized by the spherical angles $\Omega=(\theta,\phi)$ of
the orientation axis and the two parameters, $P_1^0$ for the vector and
$P_2^0$ for the tensor polarization. Furthermore, the general orientation
parameters are determined by these paramters $\Omega=(\theta,\phi)$, 
$P_1^0$ and $P_2^0$, i.e.\
\beqa
P_{IM}({\Omega})=P_I^0\,D^I_{0M}(0,-{ \theta},-{
\phi})=P_I^0e^{iM{ \phi}}d^I_{M0}({ \theta})\quad(I=1,2)\,,
\eeqa
where $D^I_{0M}$ denotes a rotation matrix~\cite{Ros57,Edm57} describing the
transformation of irreducible tensors under a spatial rotation.
Correspondingly, the density matrix with $|1m\rangle$ as basis states becomes
\beqa
\fbox{$\displaystyle
\rho^{(1)}_{m'm}=\frac{(-)^{1-m}}{\sqrt{3}}\,\sum_{I=0}^2{ P_I^0}
\sum_{M=-I}^I \widehat{I}
\left(\begin{matrix}
1 & 1 & I\cr
m & -m' & -M\cr
\end{matrix}\right)e^{iM{ \phi}}d^I_{M0}(\theta)\,,$}\label{spin_1}
\eeqa
where for convenience I have introduced $P_0^0=1$. The symbol in round
brackets denotes a 3j-symbol~\cite{Edm57}. 

\subsection{Generalization to Massive Particles with Arbitrary Spin}
The foregoing density for a massive spin-1 particle can easily be generalized
to massive particles with arbitrary spins
\beqa
\rho^{(J)}=\frac{1}{{\widehat J}^2}\,
\Big(\mathbb{1}_{2J+1}+{\widehat J}\sum_{I=1}^{2J}\sum_{M=-I}^I 
(-)^M \Omega^{[I]}_{-M}P_{IM}\Big),
\eeqa
with the reduced matrix elements of the irreducible operators $\Omega^{[I]}$
\beqa
\langle J||\Omega^{[I]}||J\rangle=\widehat J\widehat I\,.
\eeqa
This then gives the density matrix for a massive spin-J particle in the form
\beqa
\fbox{$\displaystyle
\rho^{(J)}_{m'm}=\frac{(-)^{J-m}}{\widehat J}\,\sum_{I=0}^{2J}
\widehat I\sum_{M=-I}^I
\left(\begin{matrix}
J & J & I\cr
m & -m' & -M\cr
\end{matrix}
\right)P_{IM}\,.$}
\eeqa
The associated orientation parameters $P_{IM}$ are obtained by inversion
\beqa
P_{IM}=\widehat J\widehat I\sum_{m',m}(-)^{J-m}
\left(\begin{matrix}
J & J & I\cr
m & -m' & -M\cr\end{matrix}
\right)\rho^{(J)}_{m'm}\,.
\eeqa
I will now assume again an orientation axis at {$\Omega=(\theta,\phi)$}, for
which $\rho^{(J)}$ becomes diagonal ($\rho^{(J)}_{m'm}=\delta_{m'm}p_m$) with
$2J$ orientation parameters $P_I^0$ ($I=1,\dots,2J$). These are related to the
diagonal elements $p_m$ by
\beqa
P_I^0=\widehat J\widehat I\sum_{m}(-)^{J-m}
\left(\begin{matrix}
J & J & I\cr
m & -m & 0\cr\end{matrix}
\right)p_m\,.\label{orient_par}
\eeqa
Then, using the relation
\beqa
P_{IM}({\Omega})=P_I^0\,D^I_{0M}(0,-{\theta},-{
\phi})=P_I^0e^{iM{\phi}}d^I_{M0}({\theta}), 
\eeqa
the density matrix becomes finally as a generalization of eq.~(\ref{spin_1})
\beqa
\fbox{$\displaystyle
\rho^{(J)}_{m'm}=\frac{(-)^{J-m}}{\widehat J}\,\sum_{I=0}^{2J}{ P_I^0}
\sum_{M=-I}^I\widehat I
\left(\begin{matrix}
J & J & I\cr
m & -m' & -M\cr\end{matrix}
\right)e^{iM{\phi}}d^I_{M0}({\theta})\,.$}
\eeqa

%% file: proc_em_reactions.tex
\section{Electromagnetic Reactions }
I will now turn to illustrate the use of polarization observables in photo
reactions and electron scattering on nuclei. An extensive discussion of
polarization effects in electron scattering can be found in Ref.~\cite{Don91}.

\subsection{Nuclear photo reactions}
The cross section of photo absorption on a nucleus $N$ 
\beqa
\gamma(k)+N(p_i)\rightarrow N'(p_f)\,,
\eeqa
with an initial density matrix $\rho^{(N)}_i$, characterizing the target
preparation, and a final density matrix $\rho^{(N)}_f$, characterizing the
detector, is in first order perturbation given by
\beqa
d\sigma&=&
\frac{(2\pi)^2\alpha}{q}\,\mbox{Tr}\Big(d{\cal W}(\rho^{(N)}_f,\rho^{(N)}_i)\,,
\rho^{(\gamma)}\Big),
\eeqa
where $\alpha$ denotes the fine structure constant, $q$ the photon momentum and
$\rho^{(\gamma)}$ the photon density matrix~\cite{FoW66,EiG70}. The trace
refers here to the photon polarization indices. Furthermore, $d{\cal W}$
denotes the hadron electromagnetic current tensor which depends on the density
matricesof the nuclear initial and final states, i.e.\ 
\beqa
d{\cal W}(\rho^{(N)}_f,\rho^{(N)}_i)=
\int\!\!\!\!\!\!\!\!\sum_f d^3p_f\delta^{(4)}(p_f-p_i-q)
\mbox{Tr}({\cal J}_{fi}^\dagger \rho^{(N)}_f{\cal J}_{fi}\,\rho^{(N)}_i)\,,
\eeqa
with ${\cal J}_{fi}=(j_{fi,1},j_{fi,-1})$ as the transverse nuclear current in
a spherical basis with $z$-axis along $\mathbf{q}$ for $\lambda=\pm 1$
\beqa
j_{fi,\lambda}&=&\langle \mathbf p_f|j_\lambda(0)|\mathbf p_i\rangle
=\frac{1}{(2\pi)^3}
\langle f|\widetilde j_\lambda(\mathbf q,\mathbf P)|i\rangle\,.
\eeqa
Here $\widetilde j_\lambda(\mathbf q,\mathbf P)$ denotes the {\sc Fourier}
transform of the nuclear current operator with $\mathbf q=\mathbf p_f-\mathbf
p_i$ and $\mathbf P=\mathbf p_f+\mathbf p_i$, and $|i\rangle$ and $|f\rangle$
the intial and final intrinsic nuclear states, respectively.
Evaluation of the trace with respect to the photon indices leads to
the well-known form
\beqa
d\sigma&=&\frac{4\pi^2\alpha}{q}\,
(dF_T+P^\gamma_ldF_{TT}+P^\gamma_cdF_{T}')\,,
\eeqa
where the nuclear current structure functions are defined in terms of the
hadron current tensor by
\beqa
dF_T=\frac{1}{2}\sum_{\lambda=\pm 1}d{\cal W}_{\lambda\lambda},\quad
dF_{TT}=\Re e(d{\cal W}_{-11}),\quad
dF_{T}'=\frac{1}{2}\sum_{\lambda=\pm 1}\lambda\,d{\cal W}_{\lambda\lambda}\,.
\eeqa
As noted above, the tensor $d{\cal W}=d{\cal W}(\rho^{(N)}_f,\rho^{(N)}_i)$
still depends on the initial and final nuclear density matrices
\beqa
d{\cal W}_{\lambda'\lambda}&=&
\int\!\!\!\!\!\!\!\!\sum_f d^3p_f\delta^{(4)}(p_f-p_i-q)
\mbox{Tr}({\cal J}_{fi,\lambda'}^\dagger\rho^{(N)}_f
{\cal J}_{fi,\lambda}\rho^{(N)}_i)\,, 
\eeqa
where here the trace refers to the nuclear initial and final state spin
degrees of freedom. 

For unpolarized nuclei ($\rho^{(N)}_i=\mathbb{1}_{2J_i+1}/\widehat J_i^2$) and
no polarization analysis of the final states
($\rho^{(N)}_f=\mathbb{1}_{2J_f+1}$), there cannot be any dependence on
circular photon polarization and thus the corresponding structure function
vanishes in this case ($dF_{T}^{\prime 0}=0$). The remaining structure
functions $dF_T^0$ and $dF_{TT}^0$ determine the cross section
\beqa
\fbox{$\displaystyle
d\sigma=d\sigma^0(1+P^\gamma_l\Sigma)\,,$}
\eeqa
with the unpolarized cross section $d\sigma^0$ and the linear photon asymmetry
$\Sigma$
\beqa
d\sigma^0=\frac{4\pi^2\alpha}{q}dF_T^0,\quad \Sigma=\frac{dF_{TT}^0}{dF_T^0}\,.
\eeqa
For the further analysis it useful to apply a multipole decomposition of the
current in order to exploit the conservation of angular momentum in
conjunction with angular momentum algebra
\beqa
\widetilde j(\mathbf q)_\lambda=-\sqrt{2\pi}\sum_{LM} i^L\widehat L\Big(
{ {\cal T}_{LM}^{(e)}}+\lambda { {\cal
T}_{LM}^{(m)}}\Big)D^L_{M\lambda}(-\phi_q,-\theta_q,\phi_q)\,, 
\eeqa
with electric ${\cal T}_{LM}^{(e)}$ and magnetic ${\cal T}_{LM}^{(m)}$
multipole operators, and $D^L_{M\lambda}$ denotes a rotation
matrix~\cite{Ros57}. 

\subsection{Total Photoabsorption by an oriented nucleus}
\begin{figure}[h]
\centerline {\includegraphics[scale=.5]{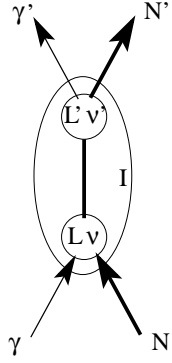}}
\caption{Total angular momentum transfer$I$ in the two-step scattering process
  induced by the multipoles $T^{\nu}_{L}$ and $T^{\nu'}_{L'}$.}
\label{polarizability}
\end{figure}

As a specific application I will now consider the total photabsorption cross
section of an oriented nucleus~\cite{Are67}. In this case one sums over all
possible final spin states ($\rho^{(N)}_f=\mathbb{1}_{2J_f+1}$) and integrates
over all possible final continuum states, like e.g.\ emitted
particles. Furthermore, according to the {\it optical theorem} the total cross
section can be related to the imaginary part of the forward elastic photon
scattering amplitude $T^C(q)$. In particular, for an oriented nucleus with
spin $J$ and density matrix $\rho^{(N)}_i=\rho^{(J)}$ one finds
\beqa
\sigma_{\mathrm{tot}}(q,\rho^{(\gamma)},\rho^{(J)})=\frac{4\pi}{q}
\Im m\Big[\mbox{Tr}\Big(\rho^{(\gamma)}\rho^{(J)}T^C(q)\Big)\Big].
\eeqa
The forward elastic scattering amplitude can be parametrized
as~\cite{ArG69,Are86} 
\beqa
T^C_{\lambda'M',\lambda M}(q)=
(-)^{J-M}\,\hat J\sum_{I=0}^{2J}\hat I\,
\left(\begin{matrix}
J & I & J\cr -M' & \lambda - \lambda' & M\cr\end{matrix}\right)
{ T^I_{\lambda'\lambda}(q)}\,,
\eeqa
where the amplitude $T^I_{\lambda'\lambda}(q)$ describes the contribution of
the total angular momentum transfer $I$ to the ground state in the scattering
process. It is given in terms of generalized polarizabilities 
$P^{L'L\lambda'\lambda}_I$~\cite{ArG69,Are86}
\begin{eqnarray}
{ T^I_{\lambda'\lambda}(q)}=\frac{\hat I}{\hat J}\,\sum_{L'L}(-)^{L'+L}
\left(\begin{matrix}
L & L' & I\cr \lambda & - \lambda' & \lambda'-\lambda\cr
\end{matrix}\right)
{ P^{L'L\lambda'\lambda}_I(q)}\,,\label{TIlambda}
\end{eqnarray}
which contain the contributions of the multipoles $T^{\nu}_{L}$ and
$T^{\nu'}_{L'}$ in the two-step scattering process, and are defined by
\begin{eqnarray}
{ P^{L'L\lambda'\lambda}_I(q)}=\sum_{\nu',\nu=0,1}\lambda^{\prime\nu'}
\,\lambda^{\nu}{ P_I(T^{\nu'}_{L'},T^\nu_L;k)}\,,
\end{eqnarray}
with the notation: $T^0_L=E_L$ for the electric and $T^1_L=M_L$ for the
magnetic multipoles. The generalized polarizabilities describe the total
angular momentum transfer $I$ in the two-step photon scattering process
induced by the two multipoles $T^\nu_L$ and $T^{\nu'}_{L'}$, which in turn
transfer angular momenta $L$ and $L'$, respectively (see
Fig.~\ref{polarizability}). 
Using the explicit expressions for $\rho^{(\gamma)}$ and $\rho^{(J)}$ one
obtains applying angular momentum algebra
\begin{eqnarray}
{ \sigma_{\mathrm{tot}}}(q,\rho^{(\gamma)},\rho^{(J)})
&=&\sigma_0^{11}(q)+\frac{1}{2}\,
\sum_{I=1}^{2J}{ P_I^0}\,\Big[(1+(-)^I)\,{ \sigma_I^{11}(q)}
+(1-(-)^I)\,{ P_c^\gamma}\,{ \sigma_I^{11}(q)}\,
P_I(\cos\theta)\nonumber\\
&&\hspace*{2cm}+(1+(-)^I){ P_l^\gamma}\,{ \sigma_I^{-11}(q)}
\,d_{20}^I({ \theta})\,\cos(2{ \phi})\Big]\,,
\end{eqnarray}
where $|P_l^\gamma|$ and $|P_c^\gamma|$ denote the degree of linear and
circular photon polarization, respectively, $P_I^0$ the nuclear orientation
parameters, and $P_I(\cos\theta)$ a {\sc Legendre} polynomial. The direction
of the 
orientation axis is determined by the angles $\theta$ and $\phi$. Furthermore,
the various contributions to the total cross section are related to the
$T^I_{\lambda'\lambda}$ (see eq.~(\ref{TIlambda})) by
\begin{eqnarray}
{ \sigma_I^{\lambda'\lambda}(q)}=
\frac{4\pi}{q}\,\Im m\,(T^I_{\lambda'\lambda}(q))\,.
\end{eqnarray}
Variation of the photon polarization parameters $P^\gamma_l$, $P^\gamma_c$,
$\theta$ and $\phi$ and the nuclear polarization parameters $P^0_I$ allows one
to determine all contributions separately. 

Of particular interest is the so-called spin asymmetry of the total cross
section defined as the difference of cross sections for right and left
circularly polarized photons on a target with maximal spin projection on the
photon momentum, i.e.\ with orientation parameters (see eq.~(\ref{orient_par})
with $p_m=\delta_{mJ}$)
\beqa
P^0_I=\widehat J\widehat I
\left(\begin{matrix}
J & J & I\cr
J & -J & 0\cr\end{matrix}
\right).
\eeqa
Then one finds for the spin asymmetry
\beqa
\sigma_{\mathrm{tot}}^P(q)-\sigma_{\mathrm{tot}}^A(q)
&=&\widehat J\sum_{I=1}^{2I}\widehat I
\left(\begin{matrix}
J & J & I\cr
J & -J & 0\cr\end{matrix}
\right)(1-(-)^I)\,\sigma_I^{11}(q)\nonumber\\
&=&\frac{2\sqrt{3J}}{\sqrt{J+1}}\sigma_1^{11}(q)+\cdots
\eeqa
This asymmetry determines the {\sc Gerasimov-Drell-Hearn} sum
rule~\cite{GDH,Are01} 
\beqa
\fbox{$\displaystyle
\int_0^\infty \frac{dq}{q}\Big(\sigma^P(q)-\sigma^A(q)\Big)
\,=\,\, 4\,\pi^2 \kappa^2\frac{e^2}{M_N^2}\,J\,,$}
\eeqa
where $M_N$ denotes the nuclear mass and $\kappa$ the anomalous magnetic
moment  of the nuclear ground state. This sum rule thus links a ground state
property, $\kappa$, to the whole internal excitation spectrum.

\subsection{Deuteron photodisintegration}
As next application I will discuss the photodisintegration of the deuteron.
The differential cross section for partially polarized photons and oriented
deuterons is~\cite{ArS91}
\beqa
\frac{d\sigma(\theta)}{d\Omega}
&=&\mbox{Tr}((T^{\gamma})^{\dagger}T^{\gamma}\,\rho^{(\gamma)}\rho^{(d)})
\nonumber\\
&=&\frac{d\sigma_0}{d\Omega}\Big(1+P^\gamma_l{ \Sigma^l(\theta)}\cos2\phi
\nonumber\\&&
+\sum_{I=1,2}{ P^d_I}\Big\{\sum_{M\leq 0}\Big({
T_{IM}(\theta)}\cos{(M({ \phi_d}-\phi) 
-\delta_{I1}\frac{\pi}{2})}
+P^\gamma_c{ T^c_{IM}(\theta)}
\sin{(M({ \phi_d}-\phi)+\delta_{I1}\frac{\pi}{2})}
\Big)d^I_{M0}({ \theta_d})\nonumber\\
&&\hspace*{3cm}+P^\gamma_l\sum_{M=-I}^I{ T^l_{IM}(\theta)}\cos{(M({
\phi_d}-\phi)+2\phi 
-\delta_{I1}\frac{\pi}{2})}d^I_{M0}({ \theta_d})\Big\}\Big).
\eeqa
This expression defines the various photon, target and photon-target
asymmetries $\Sigma^l$, $T_{IM}$, $T^c_{IM}$, and $T^l_{IM}$. 
Of particular interest is the vector target asymmetry for circularly polarized
photons $T_{10}^c$, because it determines the GDH sum rule. A few examples
should serve for illustrating the importance of polarization observables.

\subsubsection{Linear photon asymmetry for an unoriented deuteron target}
\begin{figure}[h]
\centerline{\includegraphics[scale=.7]{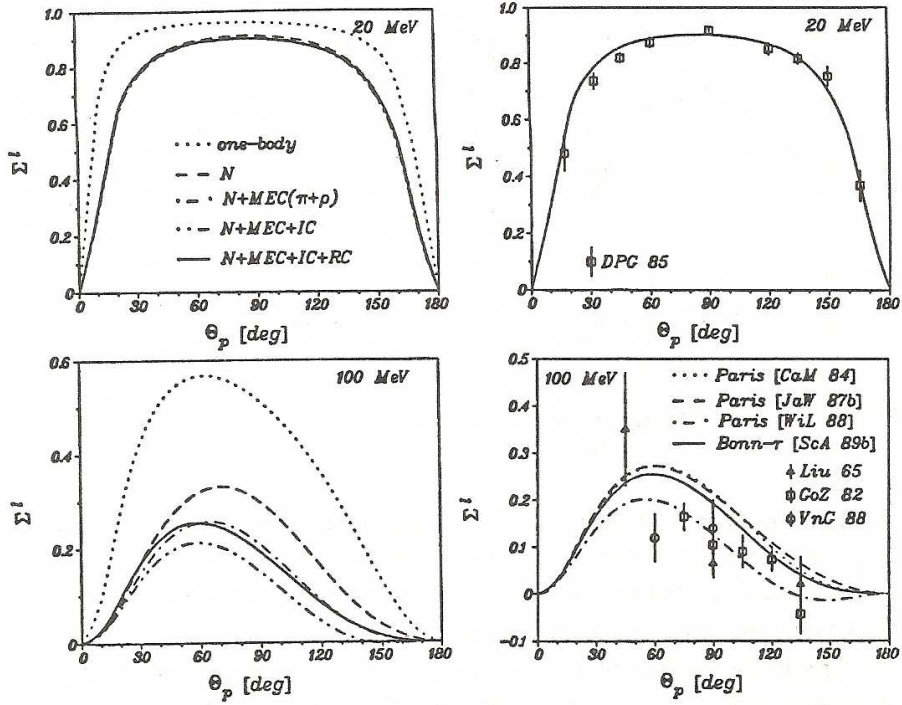}}
\caption{Photon asymmetry $\Sigma$ for deuteron photo
  disintegration at 20 and 100 MeV with contributions from meson exchange
  currents (MEC), isobar configurations (IC), relativistic effects (RC) (left
  panels), and comparison to experiments (right panels) (from~\cite{ArS91}).}
\label{asymmetry}
\end{figure}

For an unoriented deuteron, i.e.\ $P^d_1=P^d_2=0$, and photons linearly
polarized perpendicular and parallel to the reaction plane, the
corresponding cross sections are
\beqa
\frac{d\sigma_\perp}{d\Omega}&=& 
\frac{d\sigma_0}{d\Omega}(1-P^\gamma_l\Sigma(\theta))
\quad\mbox{for}\quad\phi=\pi/2\quad\mbox{(perpendicular polarization)}\,,\\
\frac{d\sigma_\|}{d\Omega}&=& 
\frac{d\sigma_0}{d\Omega}(1+P^\gamma_l\Sigma(\theta))
\quad\mbox{for}\quad\phi=0\quad\mbox{(parallel polarization)}\,.
\eeqa
Thus measuring the cross sections for parallel and perpendicular linear
polarization allows one to determine the photon asymmetry
\beqa
\fbox{$\displaystyle
\Sigma(\theta)=\frac{1}{P^\gamma_l}
\frac{d\sigma_\|-d\sigma_\perp}{d\sigma_\|+d\sigma_\perp}\,.$}
\eeqa
As is demonstrated by the example in Fig.~\ref{asymmetry}, this asymmetry
$\Sigma$ is quite sensitive to the interference of various multipoles 
and thus sensitive to various current contributions from meson exchange, isobar
configurations and relativistic effects, which enter differently into the
multipoles.

\subsubsection{Target asymmetries for unpolarized photons and
  oriented deuterons} 
\begin{figure}
\centerline{\includegraphics[scale=.6]{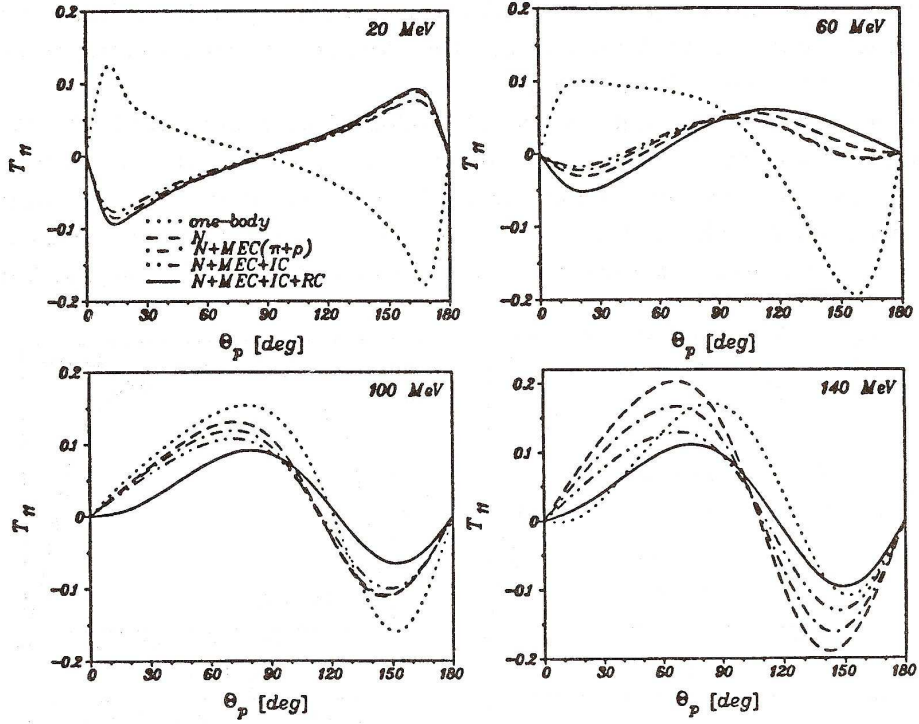}}
\caption{Vector target asymmetry $T_{11}$ for deuteron photo
disintegration with various contributions (see caption of
Fig.~\ref{asymmetry}) from~\cite{ArS91}.} 
\label{target_asy}
\end{figure}

By proper choices of the deuteron orientation axis one can extract the target
asymmetries $T_{IM}$ for unpolarized photons, namely one has
\beqa
T_{11}&=&\frac{1}{\sqrt{2}P^d_1}
\frac{d\sigma^\uparrow-d\sigma^\downarrow}{d\sigma_0},\quad
T_{20}=\frac{1}{P^d_2}\Big(2-\frac{d\sigma^\odot
+\frac{1}{2}(d\sigma^\uparrow+d\sigma^\downarrow)}{d\sigma_0}\Big),\quad
T_{22}=\frac{\sqrt{2}}{\sqrt{3}P^d_2}
\frac{d\sigma^\odot-\frac{1}{2}
(d\sigma^\uparrow+d\sigma^\downarrow)}{d\sigma_0}\,,
\eeqa
where the symbols characterize the various choices for the deuteron
orientation angles, i.e.\ perpendicular to the photon momentum
($\theta_d=90^\circ$), either in the reaction plane $(\odot)$ or 
perpendicular to the reaction plane $(\uparrow,\,\downarrow)$: 
\beqa
\odot:\,\, \phi_d=0^\circ\,,\quad \uparrow:\,\, \phi_d=90^\circ\,,\quad 
\downarrow:\,\, \phi_d=-90^\circ\,.
\eeqa
As an example, Fig.~\ref{target_asy} shows the vector target asymmetry
$T_{11}$ at various energies with 
contributions from meson exchange currents (MEC), isobar configurations (IC),
relativistic effects (RC), which again shows strong sensitivity to the
different dynamical ingredients.

\subsubsection{Spin asymmetry of the total cross section of deuteron
  photodisintegration} 
\begin{figure}
\centerline{\includegraphics[scale=.9]{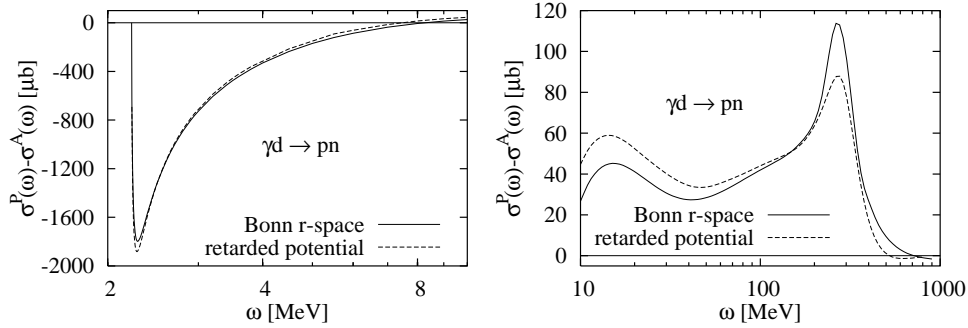}}
\caption{Spin asymmetry for deuteron photodisintegration using the realistic
  Bonn r-space $NN$-interaction and a potential which takes into account
  retardation of the exchanged pion~\cite{ArF04}.}
\label{spin_asy}
\end{figure}

\begin{figure}
\centerline{\includegraphics[scale=.9]{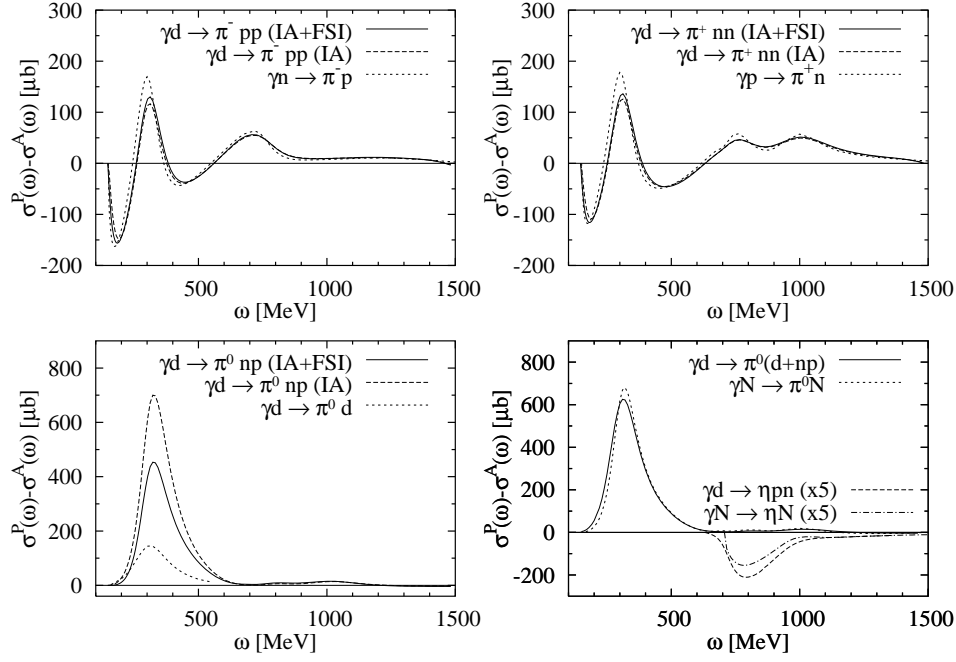}}
\caption{Spinasymmetry for pion photoproduction on deuteron and nucleon
  including pion end $\eta$-meson production~\cite{ArF04}.}
\label{pion}
\end{figure}

As already mentioned above, the vector beam-target $T_{10}^c$ asymmetry of the
total cross section of deuteron photodisintegration for circularly polarized
photons determines the spin asymmetry which enters the GDH sum rule 
\beqa
\frac{d\sigma}{d\Omega}|_{\theta_d=0,P^\gamma_c=1}-
\frac{d\sigma}{d\Omega}|_{\theta_d=0,P^\gamma_c=-1}
=2T_{10}^c \frac{d\sigma_0}{d\Omega}\,.
\eeqa
Fig.~\ref{spin_asy} shows this spin asymmetry from threshold up to 1~GeV.
It starts negativ near threshold. The reason for this feature
is easily understood. Near threshold the dominant contribution to the total
cross section arises from the $M1$ transition to the final
$^1S_0$-state. Since the total spin is zero it can only be reached when photon
and deuteron spins have opposite direction, which means a negative
$T_{10}^c$. Finally I show the spin asymmetry for pion photoproduction on
deuteron and nucleon in Fig.~\ref{pion}.

\subsection{Electron scattering and the electric form factor of the neutron}

The cross section of electron scattering on a nucleus $N$ 
\beqa
e(k_i)+N(p_i)\rightarrow e(k_f)+N'(p_f)\,,
\eeqa
with an initial density matrix $\rho^{(N)}_i$ and a final density matrix
$\rho^{(N)}_f$ has the formal appearance of the absorption of a virtual
photon~\cite{ArD74}  
\beqa
\frac{d^2\sigma}{d\epsilon_f d\Omega_e}&=&
\frac{2\alpha^2}{q_\mu^4}\frac{k_f}{k_i}\,
\int\!\!\!\!\!\!\!\!\sum_f d^3p_f\delta^{(4)}(p_f-p_i-q)
\mbox{Tr}\Big(d{\cal W}(\rho^{(N)}_f,\rho^{(N)}_i)
\rho^{(\gamma)}\Big),
\eeqa
where the nuclear current matrix elements comprise, besides the two transverse
contributions, the nuclear charge as zeroth component which has been introduced
in favour of the longitudinal one via current conservation, i.e.\ ${\cal
  J}_{fi}=(j_{fi,0},j_{fi,1},j_{fi,-1})$. Furthermore, the virtual photon
density matrix for unpolarized electrons is given as a $3\times3$-matrix
\beqa
(\rho^{(\gamma^*)0})_{\lambda'\lambda}&=&-\frac{1}{2}q_\mu^2\,
\left(\begin{matrix}
\frac{\xi^2}{\eta} &\frac{\xi}{\eta}\sqrt{\frac{1}{2}(\xi+\eta)} &  
-\frac{\xi}{\eta}\sqrt{\frac{1}{2}(\xi+\eta)}\cr 
\frac{\xi}{\eta}\sqrt{\frac{1}{2}(\xi+\eta)} & 1+\frac{\xi}{2\eta} & 
-\frac{\xi}{2\eta} \cr 
-\frac{\xi}{\eta}\sqrt{\frac{1}{2}(\xi+\eta)} &-\frac{\xi}{2\eta} & 
1+\frac{\xi}{2\eta}
\end{matrix}\right),
\eeqa
with $\lambda=0,1,-1$, $\xi=-q_\mu^2/\mathbf{q}^2$, $\eta=\tan^2(\theta_e/2)$,
$q$ as momentum transfer, and $\theta_e$ as the electron scattering angle. 
For longitudinally polarized electrons of degree $h$ the virtual photon
density matrix has to be supplemented acoording to
\beqa
\rho^{(\gamma^*)}=\rho^{(\gamma^*)0}+h\,\rho^{(\gamma^*)\prime}\,,
\eeqa
where
\beqa
(\rho^{(\gamma^*)\prime})_{\lambda'\lambda}&=&
\frac{1}{2}q_\mu^2\,\left(\begin{matrix}
 0& \frac{\xi}{\sqrt{2\eta}} & \frac{\xi}{\sqrt{2\eta}}\cr 
 \frac{\xi}{\sqrt{2\eta}} &\sqrt{1+\frac{\xi}{\eta}} & 0 \cr 
\frac{\xi}{\sqrt{2\eta}} & 0 &  -\sqrt{1+\frac{\xi}{\eta}}
\end{matrix}\right).
\eeqa
One should notice that the trace of $\rho^{(\gamma^*)}$ is not normalized.
Instead one has
\beqa
\mbox{Tr}(\rho^{(\gamma^*)})=
-\frac{1}{2}q_\mu^2\,(\frac{\xi^2}{\eta}+\frac{\xi}{\eta}+2)\,. 
\eeqa
The normalized density matrix $\widetilde\rho^{(\gamma^*)}$ is given by
\beqa
(\widetilde\rho^{(\gamma^*)})_{\lambda'\lambda}&=&
\frac{1}{\xi^2+\xi+2\eta}\,\left(\begin{matrix}
{\xi^2} &\widetilde\rho_{01} &  \widetilde\rho_{0-1}\cr 
\widetilde\rho_{01} & \eta+\frac{\xi}{2} -h\sqrt{\eta(\xi+\eta)}& 
-\frac{\xi}{2} \cr 
\widetilde\rho_{0-1} &-\frac{\xi}{2} & 
\eta+\frac{\xi}{2}+h\sqrt{\eta(\xi+\eta)}
\end{matrix}\right),
\eeqa
where for convenienc I have introduced $\widetilde\rho_{0\pm
  1}=\pm{\xi}\sqrt{\frac{1}{2}(\xi+\eta)}-h\xi\frac{\eta}{2}$.
This leads to the following interpretation of the virtual photon density
matrix: 
\begin{description}
\item{(i)}
$P_{\mathrm{long}}=\frac{\xi^2}{\xi^2+\xi+2\eta}$ describes the fraction of
longitudinal photons.
\item{(ii)}
$P_{\mathrm{trans}}=\frac{\xi+2\eta}{\xi^2+\xi+2\eta}$ describes the fraction
of transverse photons.
\item{(ii)}
The ratio longitudinal to transverse photons is
$P_{\mathrm{long}}/P_{\mathrm{trans}}= \frac{\xi^2}{\xi+2\eta}$. For
$\theta_e\to 0$ this ratio approaches
$P_{\mathrm{long}}/P_{\mathrm{trans}}\to\xi=1-\omega^2/\mathbf{q}^2$, since
then $\eta\to 0$. 
\end{description}
Furthermore, the transverse part of $\widetilde\rho^{(\gamma^*)}$ is
\beqa
\rho^{(\gamma^*)}_t&=&
\frac{\eta}{\xi^2+\xi+2\eta}\,\left(\begin{matrix}
1+\frac{\xi}{2\eta} +h\sqrt{1+\frac{\xi}{\eta}}& -\frac{\xi}{2\eta} \cr 
-\frac{\xi}{2\eta} & 1+\frac{\xi}{2\eta}-h\sqrt{1+\frac{\xi}{\eta}}
\end{matrix}\right)\nonumber\\
&=&\frac{\eta+\frac{\xi}{2}}{\xi^2+\xi+2\eta}
(\mathbb{1}_2-{\varepsilon_l}\sigma_x
+{ h\varepsilon_c}\sigma_z)\,.
\eeqa
This means that for unpolarized electrons the transverse part of the virtual
photon density matrix describes linear polarization in the scattering plane of
degree {$\varepsilon_l=\xi/(\xi+2\eta)$}. Furthermore, longitudinally
polarized electrons create in addition circular polarization of the virtual
photon of degree {$h\varepsilon_c$} with
{$\varepsilon_c=2\sqrt{\eta(\xi+\eta)}/(\xi+2\eta)$}. 

Evaluation of the trace with respect to the virtual photon polarization
indices leads to the well-known form of the differential scattering cross
section~\cite{FoW66,EiG70} 
\beqa
\frac{d^2\sigma}{d\epsilon_f d\Omega_e}&=&
\frac{2\alpha^2}{q_\mu^4}\frac{k_f}{k_i}\,
(\rho_LdF_L+\rho_TdF_T+\rho_{LT}dF_{LT}+\rho_{TT}dF_{TT}\nonumber\\
&&\hspace*{1cm}+h(\rho_{LT}'dF_{LT}'+\rho_{T}'dF_{T}'))\,,
\eeqa
with the electron kinematical functions
\beqa
\rho_L=\rho^{(\gamma^*)0}_{00},\,\, \rho_T=\rho^{(\gamma^*)0}_{11},\,\,
\rho_{LT}=\rho^{(\gamma^*)0}_{01},\,\, \rho_{TT}=\rho^{(\gamma^*)0}_{-11},\,\,
\rho_{LT}'=\rho^{(\gamma^*)\prime}_{01},\,\,
\rho_{T}'=\rho^{(\gamma^*)\prime}_{11},
\eeqa
and the nuclear current structure functions
\beqa
\begin{array}{ll}
dF_L=d{\cal W}_{00}, & 
dF_T=\sum_{\lambda=\pm 1}d{\cal W}_{\lambda\lambda},\\
 & \\
dF_{LT}=2\Re e(d{\cal W}_{01}+d{\cal W}_{10}), & 
dF_{TT}=2\Re e(d{\cal W}_{-11}),\\
 & \\
dF_{LT}'=-2\Im m(d{\cal W}_{01}+d{\cal W}_{10}), & 
dF_{T}'=\sum_{\lambda=\pm 1}\lambda d{\cal W}_{\lambda\lambda}.\\
\end{array}
\eeqa
The hadron current tensor  d${\cal W}_{\lambda'\lambda}$ still depend on the
initial and final orientation parameters 
\beqa
d{\cal W}_{\lambda'\lambda}&=&
\int\!\!\!\!\!\!\!\!\sum_f d^3p_f\delta^{(4)}(p_f-p_i-q)
\mbox{Tr}({\cal J}_{fi,\lambda'}^\dagger\rho^{(N)}_f
{\cal J}_{fi,\lambda}\rho^{(N)}_i
)\,,
\eeqa
where here the trace refers to the nuclear spin degrees of freedom. 

\subsubsection{Deuteron electrodisintegration and the neutron electric form
  factor} 

In the absence of neutron targets of sufficient density one uses light nuclei
as effective neutron targets, deuteron or $^3$He, and studies, for example,
inelastic electron scattering off the deuteron $d(e,e'n)p$ in quasi-free
kinematics, where one observes the ougoing neutron. Quasi-free means that
the emitted neutron takes up energy and momentum transfer such that the
recoiling proton remains at rest in the laboratory. 

\begin{figure}
\centerline{\includegraphics[scale=.5]{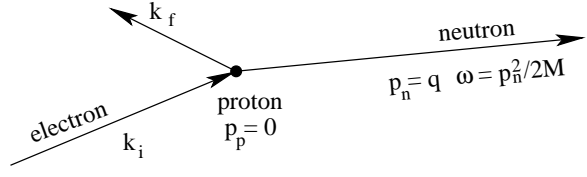}}
\caption{Kinematics of quasi-free electron scattering on the deuteron in the
  laboratory.} 
\label{quasifree}
\end{figure}

For this kinematics the scattering cross section is roughly given by the cross
section on the neutron multiplied by the probability of finding a neutron at
rest within the deuteron. In this way, the neutron magnetic form factor $G_M$ 
can be determined. 
However, since the electric form factor $G_E$ of the neutron is much
smaller than its magnetic one, its contribution is buried under competing
binding and final state interaction effects. In this situation polarization
observables come into play as they offen allow one to investigate small
contributions by interference with large ones. 

In order to illustrate this, I will first consider an exercise with a
simplified discussion of elastic electron-neutron scattering. Assuming a
nonrelativistic form of charge and transverse current~\cite{ArL88}
\beqa
j_0=G_E,\quad j_{\pm 1}=\sqrt{2}\frac{q}{2M}G_M\,\sigma_{\mp}\,,
\eeqa
and taking for the polarized initial neutron the density matrix
\beqa
\rho^{(n)}=\frac{1}{2}(\mathbb{1} +\mathbf{P}\cdot\mathbf{\sigma})\,,
\eeqa
one finds for the structure functions
\beqa
\begin{array}{lll}
F_L={G_E^2}, & \qquad F_T=\frac{q^2}{2M^2}G_M^2, &
\qquad F_{LT}=0, \\
 & & \\
F_{TT}=0, &\qquad F_{LT}'=-\sqrt{2}\frac{q}{M}G_EG_MP_x, &
\qquad F_{TT}'=-\frac{q^2}{2M}G_M^2P_z.\\
\end{array}
\eeqa
Thus one obtains as differential cross section
\beqa
\frac{d\sigma}{d\Omega_e}&=&
\frac{d\sigma_0}{d\Omega_e}(1+h\mathbf{A}\cdot\mathbf{P}),\qquad
\frac{d\sigma_0}{d\Omega_e}=c\frac{q^2}{2M^2}G_M^2\,S_0,
\eeqa
with the neutron target asymmetry
\beqa
\mathbf{A}=(-2\rho_{LT}'{ R}/S_0,0,-\rho_{T}'/S_0)\,,
\eeqa
and furthermore
\beqa
S_0=\rho_{T}+\rho_{L}{ R^2},\mbox{ and }
{ R}=\sqrt{2}\frac{M}{q}{ \frac{G_E}{G_M}}\,.
\eeqa

Thus one notes that the unpolarized cross section is dominated by the neutron
magnetic form factor $G_M$, while the electric form factor $G_E$ gives via
$R^2$ a tiny contribution only.  
But for longitudinally polarized electrons and initially polarized neutrons
with spin lying in the scattering plane and perpendicular to $\mathbf{q}$, the 
corresponding asymmetry ${A}_{x}=-2\rho_{LT}'R/S_0$ is linear in the ratio
$R$ and thus more sensitive to $G_E$. 
The same holds for an unpolarized initial neutron and the final neutron
polarization component in the scattering plane and perpendicular to
$\mathbf{q}$, i.e.\ $P_x={A}_{x}$.

\begin{figure}[h]
\centerline {\includegraphics[scale=.5]{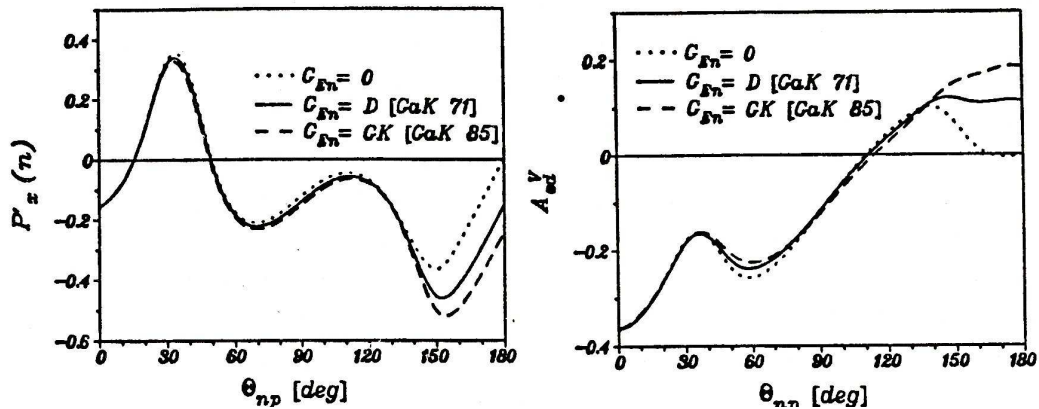}}
\caption{Two examples for the influence of the electric form factor
  of the neutron in quasi-free electron deuteron scattering at quasi-free
  kinematics ($\mathbf{q}^2=12\mbox{ fm}^{-2}$) for $G_{En}=0$ (dotted) and
  two different models for $G_{En}=0$ from~\cite{ArL88}. 
  Left panel: Polarization component $P_x'$ of the final neutron. Right panel:
  Vector asymmetry $A_{ed}^V$ at $\theta_d=90^\circ, \phi_d=0$. Dotted
  $G_{En}=0$}
\label{neutron}
\end{figure}

Now I will return to quasifree scattering on the deuteron. 
One needs a vector polarized deuteron target for having an initially polarized
neutron. The polarization component of the bound neutron in a vector polarized
deuteron along the orientation axis is given by
\beqa
{ P=(p_1-p_{-1})(1-\frac{3}{2}P_D) }, 
\eeqa
with $P_D$ for the $D$-wave probability. 
In Born approximation, neglecting final state interactions and also the
small $D$-component in the deuteron, the cross section in quasi-free
kinematics for polarized electrons and a vector-polarized deuteron target has 
the form
\beqa
\frac{d\sigma}{d\Omega_ed\Omega_{np}}&=&
\frac{d\sigma_0}{d\Omega_ed\Omega_{np}}(1+hP_1^d{ {A}^V_{ed}}),\qquad
\frac{d\sigma_0}{d\Omega_e}=c'\frac{q^2}{2M^2}G_M^2\,S_0\,,
\eeqa
where the vector asymmetry is given by
\beqa
{ {A}^V_{ed}}&=&\frac{1}{S_0}
\sqrt{\frac{2}{3}}\Big({\rho_T'}\cos\theta_d
+2{\rho_{LT}'}\sin\theta_d\cos\phi_d\,{ R}
\Big)\,.
\eeqa
Thus indeed, this vector asymmetry depends linearly on the ratio R. In
particular one finds for $\theta_d=90^\circ$ and $\phi_d=0$
\beqa
{ {A}^V_{ed}}&=&\frac{2}{S_0}
\sqrt{\frac{2}{3}}{\rho_{LT}'}\,{ R}\,,
\eeqa
which means, that for this kinematics one has direct access to the ratio
containing the electric form factor of the neutron. Instead of using an
initially polarized neutron, one can measure the polarization of the outgoing
neutron in $d(\vec{e},e'\vec{n})p$ for
which one obtains 
\beqa
{ P_x'(n)}=\frac{2}{S_0}\,{\rho_{LT}'}\,{ R}=\sqrt{\frac{3}{2}}{
{A}^V_{ed}}. 
\eeqa
An example is shown in Fig.~\ref{neutron}, where one readily notices the
strong influence of the electric form factor of the neutron right on the
quasi-free peak ($\theta_{np}=180^\circ$). Naturally, one has to check what the
influence is of the neglected effects from final state interaction,
relativistic effects and additional current contributions from meson exchange
and isobar configurations. Indeed, a thorough calculation has shown that
indeed these additional effects have quite small
effect at $\theta_{np}=180^\circ$.  

%% file: proc_complete_sets.tex
\section{Complete Sets of Observables }
The polarization observables of a reaction, i.e.\ beam and target asymmetries,
polarization components of the outgoing particles and combinations thereof,
provide a wealth of information. Thus the question naturally arises, does one
need all of them in order to obtain complete information on this reaction or
is a smaller number carefully chosen already sufficient. In order to
illustrate this problem I will consider again a reaction of the type
\begin{eqnarray}
a + b \rightarrow c + d +\cdots\,,
\end{eqnarray}
where the participating particles have spins $s_a,\,s_b,\dots$. 
Let me denote the corresponding $T$-matrix by
\begin{eqnarray}
T_{\lambda_a \lambda_b \lambda_c \dots} 
\end{eqnarray}
with helicities (spin projections on the particle momenta) $\lambda_a,\,
\lambda_b,\, \lambda_c \dots$. Angular momentum conservation requires that
$s_a+s_b$ as well as $s_c+s_d+\cdots$ are either half-integer or integer.
The whole set of these complex $T$-matrix elements contains the maximal
dynamic information available for this reaction. But in general not all
of them are independent, since there might exist relations between the
matrix elements due to some conservation laws. Therefore, it suffices to
determine the maximal subset of independent matrix elements. 
The total number of $T$-matrix elements is given by
\begin{eqnarray}
N=N(s_a) N(s_b) N(s_c) N(s_d) \cdots\,,
\end{eqnarray}
with
\begin{eqnarray}
N(s_\alpha)=\left\{
\begin{array}{ll}
2s_\alpha+1 & \quad\mbox{for massive particles}\,,\\
2           & \quad\mbox{for massless particles}\,.
\end{array}
\right.
\end{eqnarray}
If parity is conserved, the number of independent matrix elements reduces to 
\begin{eqnarray}
n=\left\{
\begin{array}{ll}
N/2 & \quad\mbox{for } N\, \mbox{even}\,,\\
(N-1)/2 & \quad\mbox{for } N\, \mbox{odd}\,.
\end{array}
\right.
\end{eqnarray}
What one measures are observables $\cal O$, like cross sections, asymmtries
etc., which are hermitean forms of the $T$-matrix elements
\begin{eqnarray}
{\cal O}_\alpha=\sum_{j'j}T_{j'}^*O_{j'j}^\alpha T_j\,.
\end{eqnarray}
Here, $j'$ and $j$ enumerate all independent $T$-matrix elements and
$O_{j'j}^\alpha$ denotes a hermitean $n\times n$ matrix associated with the
observable $\cal O$. For example, the unpolarized differential cross section
is given by 
\begin{eqnarray}
\frac{d\sigma}{d\Omega}\propto\sum_{j}T_{j}^*T_j\,,
\end{eqnarray}
or the polarisation component $P_{x}^c$ of the outgoing particle $c$
\begin{eqnarray}
P_{x}^c\frac{d\sigma}{d\Omega}\propto\sum_{j'j}T_{j'}^*(S_x^c)_{j'j}T_j\,,
\end{eqnarray}
with $S_x^c$ as the corresponding spin operator. 

The number of {\em linearly} independent observables is equal to the number of
linearly independent hermitean $n\times n$ matrices, namely $n^2$. 
On the other hand, the $n$ independent complex matrix elements are determined
by $2n-1$ real numbers, since a common phase remains unobserved.
This seeming contradiction is resolved by the observation that exactly
$(n-1)^2$ quadratic relations exist between the linearly independent
observables (see Appendix B of~\cite{ArL00}). Thus one can conclude
\vspace*{.3cm}

\centerline{\fbox{\parbox{11cm}{\em A set of $2n-1$ properly chosen observables
should allow a determination of all independent $T$-matrix elements except for
possible discrete ambiguities, whose elimination requires 
 additional observables.}}} 
\vspace*{.3cm}

For illustration, take the simplest case of the absorption of a scalar
particle on a $s=1/2$ particle. One has four matrix elements which reduce to
two independent complex matrix elements $T_1$ and $T_2$ if parity is
conserved. In this case, each observable is represented by a $2\times 2$
matrix for which we may choose the unit matrix $(\sigma_0=\mathbb{1}_2)$ and
the three {\sc Pauli} matrices $(\sigma_i,\, i=1,2,3)$. Thus we have four
independent observables 
\begin{eqnarray}
{\cal O}_\alpha=T^\dagger \sigma_\alpha T\,,\quad (\alpha=0,\dots,3)\,,
\end{eqnarray}
or in detail
\begin{eqnarray}
\begin{array}{ll}
{\cal O}_0=|T_1|^2+|T_2|^2\,,&
{\cal O}_1=2\,\Re e(T_1^*T_2)\,,\\
 & \\
{\cal O}_2=2\,\Im m(T_1^*T_2)\,,&
{\cal O}_3=|T_1|^2-|T_2|^2\,.\\
\end{array}
\end{eqnarray}
One easily finds the following quadratic relation
\begin{eqnarray}
{\cal O}_3^2 &=& {\cal O}_0^2 - 4|T_1|^2|T_2|^2
             = {\cal O}_0^2 - ({\cal O}_1^2+{\cal O}_2^2)\,,
\end{eqnarray}
which leads to the inequality
\begin{eqnarray}
{\cal O}_0^2&\ge&{\cal O}_1^2+{\cal O}_2^2\,.
\end{eqnarray}
Without loss of generality we can choose the overall phase such that $T_1$
becomes real and non-negative. Then the two matrix elements can be obtained
from three observables, for example
${\cal O}_0$, ${\cal O}_1$ and ${\cal O}_2$. 
\begin{eqnarray}
T_1^2&=&\frac{1}{2}\Big({\cal O}_0\pm\sqrt{{\cal O}_0^2 -({\cal O}_1^2+
        {\cal O}_2^2)}\,\Big)\,,\label{eqt1}\\
T_2 &=& \frac{1}{2T_1}({\cal O}_1+i{\cal O}_2)\,.
\end{eqnarray}
Obviously, one finds in general two solutions. 
This discrete ambiguity can be resolved by
considering in addition ${\cal O}_3$, for which one finds, inserting the
general solution,
\begin{eqnarray}
{\cal O}_3&=&\pm\sqrt{{\cal O}_0^2 -({\cal O}_1^2+{\cal O}_2^2)}
\,,
\end{eqnarray}
Thus the sign of ${\cal O}_3$ selects the proper solution.

An alternative method exploits the fact that all bilinear forms $T_i^*T_j$ can
be given as linear expressions in the observables
\begin{eqnarray}
T_1^*T_1 &=& \frac{1}{2}({\cal O}_0+{\cal O}_3)\,,\quad
T_1^*T_2 = \frac{1}{2}({\cal O}_1+i{\cal O}_2)\,,\quad
T_2^*T_2 = \frac{1}{2}({\cal O}_0-{\cal O}_3)\,.
\end{eqnarray}
Setting $T_1$ real and positive, one finds
\begin{eqnarray}
T_1&=&\sqrt{\frac{1}{2}({\cal O}_0+{\cal O}_3)}\,,\quad
T_2=\frac{{\cal O}_1+i{\cal O}_2}{\sqrt{2({\cal O}_0+{\cal O}_3)}}\,.
\end{eqnarray}
The disadvantage of this approach is that the choice of observables is fixed
by the bilinear expressions. Thus the essential question is:
\vspace*{.3cm}

\centerline{\fbox{\parbox{12cm}{\em
 Is there a criterion which allows one to
 decide whether a chosen set of $2n-1$ observables
 is complete, i.e.\ allows the determination of
 all independent $T$-matrix elements, except for
 eventual discrete ambiguities?
}}}
\vspace*{.3cm}
Indeed, such a criterion exists and has been applied successfully. For details
see~\cite{ArL98}.

\subsection{Explicit inversion}
Now I will briefly describe an alternative method based on the explicit
inversion of the observables with respect to the bilinear forms in the
$T$-matrix elements. 
As an example, I will consider as observables the longitudinal structure
functions $f_{L,X}^{IM}$ (abbreviated form $X^{IM}$) of deuteron
electro\-disintegration (for details see \cite{ArL00}). 
Here $X=x_{\alpha_1}x_{\alpha_2}$ ($\alpha_i=0,\dots,3$) describes
the polarization components $P_{\alpha_1}P_{\alpha_2}$ of the two 
outgoing nucleons ($\alpha_i=0$ means no polarization of the 
$i$-th nucleon). In particular, $X=x_0 x_0=1$ refers to the 
differen\-tial cross section. 
Furthermore, $IM$ ($0\le M\le I$) describes the deuteron polarization, i.e.\
\begin{eqnarray}
I=\left\{\begin{matrix}
0 & \mbox{ unpolarized,}\cr
1 & \mbox{ vector polarized,}\cr
2 & \mbox{ tensor polarized.} \cr
\end{matrix}\right.
\end{eqnarray}
The initial state has three spin states for the deuteron and only one for the
longitudinal virtual photon, and the final state has for each nucleon two spin
states. Thus one has 12 $T$-matrix elements which however are not independent
if parity is conserved which we will assume. Then the number reduces to
six independent matrix elements which leads to 36 linearly independent
observables. On the other hand, a complete set should comprise eleven
observables. Because of the fact that the spin operators form a complete basis
in spin space, it is possible to represent all bilinear forms $T_{j'}^*T_j$ as
linear combinations of structure functions $f^{I M}_{L,X}$ 
\begin{eqnarray}
T_{j'}^*T_j={\cal T}_{j'j}[f^{I M}_{L,X}]\,.
\end{eqnarray}
For example one finds for ${\cal T}_{j,1}=T_{j}^*T_1$ ($j=2,\dots,6$)
\begin{eqnarray}
{\cal T}_{2,1}&=& \frac{1}{6}\Big(-f_{L,\,zz}^{0 0} + 
{\sqrt{2}}\,f_{L,\,zz}^{2 0}+
  i\,( f_{L,\,xz}^{0 0} - {\sqrt{2}}\,f_{L,\,xz}^{2 0} ) \Big) \,,\\
{\cal T}_{3,1}&=& \frac{1}{4\sqrt{6}}\Big(-f_{L}^{2 1} - f_{L,\,y_1}^{2 1}+
  i\,( -f_{L}^{1 1} - f_{L,\,y_1}^{1 1} ) \Big) \,,\\
{\cal T}_{4,1}&=& \frac{1}{4\sqrt{6}}\Big(-f_{L,\,x_2}^{1 1} + 
f_{L,\,z_2}^{2 1}+
  i\,( -f_{L,\,x_2}^{2 1} - f_{L,\,z_2}^{1 1} ) \Big) \,,\\
{\cal T}_{5,1}&=& \frac{1}{4\sqrt{6}}\Big(-f_{L,\,x_1}^{1 1} + 
f_{L,\,z_1}^{2 1}+
  i\,( -f_{L,\,x_1}^{2 1} - f_{L,\,z_1}^{1 1} ) \Big) \,,\\
{\cal T}_{6,1}&=& \frac{1}{4\sqrt{6}}\Big(f_{L,\,xz}^{1 1} + f_{L,\,zz}^{2 1}+
  i\,( -f_{L,\,xz}^{2 1} + f_{L,\,zz}^{1 1} ) \Big) \,,
\end{eqnarray}
which obviously are determined by 20 observables.
These expressions can be used to relate all matrix elements to one
\begin{eqnarray}
T_i=\frac{1}{T_j^*}\,{\cal T}_{j i}\quad\mbox{for } i\ne j\,,
\end{eqnarray}
where the ${\cal T}_{j i}$ on the right hand side are given in terms of the
observables $f^{I M}_{L,X}$. The missing matrix element $T_j$ can be
determined from a properly chosen additional observable. The disadvantage of
this approach is that one needs more than 11 observables, for example, in the
above case 21. 

\begin{figure}[h]
\centerline {\includegraphics[scale=.5]{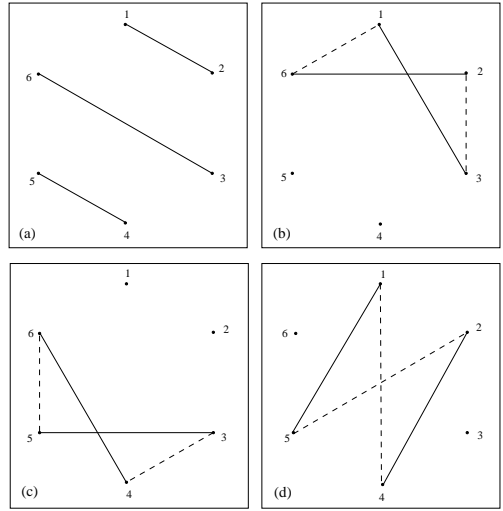}}
\caption{Diagrammatic representation of groups of longitudinal observables
  determining the interference terms of $T$-matrix elements. The nomenclature
  for the groups and the corresponding observables are listed in
  Table~\ref{table_long_hel4} (from~\cite{ArL00}).}
\label{long_hel4}
\end{figure}

One is lead to a {\em better strategy} by the observation that disjunct 
groups of observables exist, which determine a subgroup of interference 
terms, e.g.\
\begin{eqnarray}
{\cal T}_{3,1}&=& \frac{1}{4\sqrt{6}}\Big(-f_{L}^{2 1} - f_{L,\,y_1}^{2 1}+
  i\,( -f_{L}^{1 1} - f_{L,\,y_1}^{1 1} ) \Big) \,,\\
{\cal T}_{6,2}&=& \frac{1}{4\sqrt{6}}\Big(-f_{L}^{2 1} + f_{L,\,y_1}^{2 1}+
  i\,( -f_{L}^{1 1} + f_{L,\,y_1}^{1 1} ) \Big) \,,
\end{eqnarray}
or
\begin{eqnarray}
{\cal T}_{6,1}&=& \frac{1}{4\sqrt{6}}\Big(f_{L,\,xz}^{1 1} + f_{L,\,zz}^{2 1}+
  i\,( -f_{L,\,xz}^{2 1} + f_{L,\,zz}^{1 1} ) \Big) \,,\\
{\cal T}_{3,2}&=& \frac{1}{4\sqrt{6}}\Big(-f_{L,\,xz}^{1 1} + f_{L,\,zz}^{2 1}+
  i\,( f_{L,\,xz}^{2 1} + f_{L,\,zz}^{1 1} ) \Big) \,.
\end{eqnarray}
For a survey it is useful to introduce a diagrammatic representation in the
following way: Each matrix element $T_j$ is represented by a point on a circle
with equal distance between neighbouring points, and an interference term
${\cal T}_{ij}$ is represented by a straight line connecting point ``i'' with
point ``j''. Those interference terms determined by the same group of
observables are represented by the same type of lines. An example is shown in
Fig.~\ref{long_hel4}. 

\vspace*{-.5cm}
\begin{table}[h]
\caption{Nomenclature of the diagrammatic representation of groups of
  observables which determine the interference terms ${\cal T}_{ij}$ in
  Fig.~\ref{long_hel4}.} 
\renewcommand{\arraystretch}{1.2}
\begin{tabular}{|c|l|c|l|}
\hline
 Line type & Observable group & Line type & Observable group
\\
\hline
&&&\\[-2.ex]
{ (a)}  solid & $xz^{00}$, $xz^{20}$, $xz^{22}$, $zz^{00}$, &
(b) solid & $1^{11}$, $1^{21}$, $y_1^{11}$, $y_1^{21}$\\ 
   & $zz^{20}$, $zz^{22}$ & 
dashed & $xz^{11}$, $xz^{21}$, $zz^{11}$, $zz^{21}$ \\
\hline
&&&\\[-2.ex]
{ (c)} solid & $x_1^{10}$, $x_1^{22}$, $z_1^{10}$, $z_1^{22}$ & 
(d) solid & $x_1^{11}$, $x_1^{21}$, $z_1^{11}$, $z_1^{21}$ \\
     dashed & $x_2^{10}$, $x_2^{22}$, $z_2^{10}$, $z_2^{22}$ &
     dashed & $x_2^{11}$, $x_2^{21}$, $z_2^{11}$, $z_2^{21}$ 
\\[1.ex]
\hline
\end{tabular}\label{table_long_hel4}
\end{table}

\begin{figure}
\centerline {\includegraphics[scale=.5]{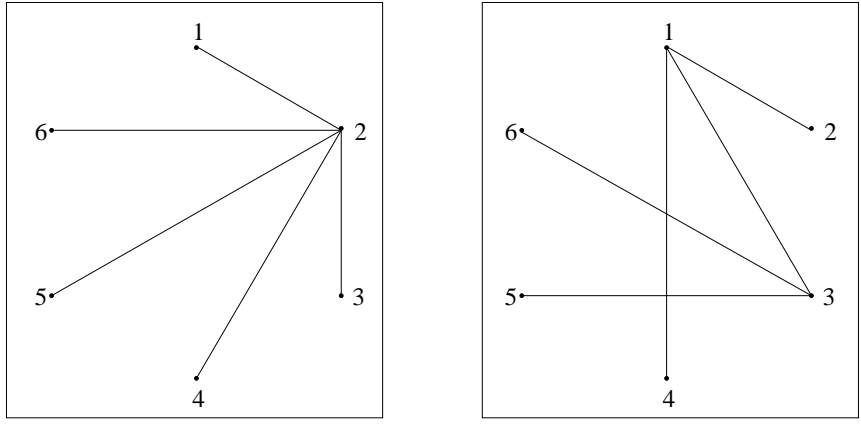}}
\caption{Examples of sets of independent interference terms}
\label{ind_set}
\end{figure}

Now, the better strategy consists in finding a set of $n-1$ {\em independent} 
interference terms. Independent means a representation by a diagram with all
$n$ points connected by exactly $n-1$ lines such that each point is
connected to all the others, not necessarily directly. 
Then closed loops cannot appear. Examples are shown in Fig.~\ref{ind_set}.

In such a set, all matrix elements can be expressed by one, because 
one finds for a matrix element $j'$, which is connected via the points  
$j_1$ through $j_k$ with point $j$,  
\begin{eqnarray}
T_{j'}=\left\{\begin{array}{ll} \frac{{\cal T}_{j_1 j'}}{{\cal T}_{j_1 j_2}}
\frac{{\cal T}_{j_3 j_2}}{{\cal T}_{j_3 j_4}}\cdots 
\frac{{\cal T}_{j_{k-2} j_{k-3}}}{{\cal T}_{j_{k-2} j_{k-1}}}
\frac{{\cal T}_{j_k j_{k-1}}}{{\cal T}_{j_k j}}\,T_j\,, & \,\mbox{if }k\,
\mbox{ odd}\\
&\\
\frac{{\cal T}_{j_1 j'}}{{\cal T}_{j_1 j_2}}
\frac{{\cal T}_{j_3 j_2}}{{\cal T}_{j_3 j_4}}\cdots 
\frac{{\cal T}_{j_{k-3} j_{k-4}}}{{\cal T}_{j_{k-3} j_{k-2}}}
\frac{{\cal T}_{j_{k-1} j_{k-2}}}{{\cal T}_{j_{k-1} j_k}}
\frac{{\cal T}_{j j_k}}{T_j^*}\,, 
& \,\mbox{if }k\,\mbox{ even}\end{array}\right.\label{loops}
\end{eqnarray}
The procedure is to select a {\em minimal number} of groups of 
obser\-va\-bles which contain an independent set of interference terms. These,
however, may in general contain also additional interference terms leading to
closed loops which constitute additional conditions. For example, 
for a loop through an even number of points, one obtains from (\ref{loops})
with $k$ odd and $j'=j$
\begin{eqnarray}
\frac{{\cal T}_{j_1 j}}{{\cal T}_{j_1 j_2}}
\frac{{\cal T}_{j_3 j_2}}{{\cal T}_{j_3 j_4}}\cdots 
\frac{{\cal T}_{j_{k-2} j_{k-3}}}{{\cal T}_{j_{k-2} j_{k-1}}}
\frac{{\cal T}_{j_k j_{k-1}}}{{\cal T}_{j_k j}}=1\,,
\end{eqnarray}
which constitutes two real quadratic relations between the observables. For a
loop with an odd number of points follows from (\ref{loops}) with $k$ even
\begin{eqnarray}
|T_{j}|^2=\frac{{\cal T}_{j_1 j}}{{\cal T}_{j_1 j_2}}
\frac{{\cal T}_{j_3 j_2}}{{\cal T}_{j_3 j_4}}\cdots 
\frac{{\cal T}_{j_{k-3} j_{k-4}}}{{\cal T}_{j_{k-3} j_{k-2}}}
\frac{{\cal T}_{j_{k-1} j_{k-2}}}{{\cal T}_{j_{k-1} j_k}}\,{\cal T}_{j j_k}\,,
\end{eqnarray}
by which the moduli of all participating matrix elements are fixed. 
Examples of {\em minimal sets} are shown in Fig.~12.
\begin{figure}
\centerline {\includegraphics[scale=.5]{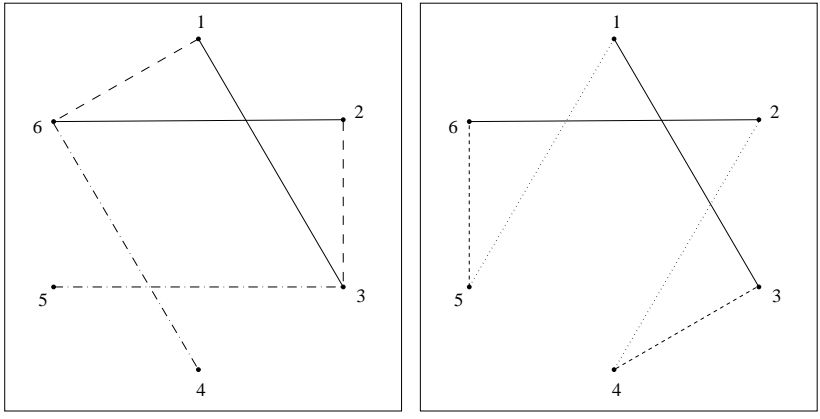}}
\caption{Examples of {\em minimal sets} containig a 4-loop (left) and a 6-loop
(right)}
\end{figure}
In both cases three groups of observables are participating, which 
each consists of four observables. Because of the loop through an even 
number of points, one can eliminate two of the observables. Thus all 
matrix elements can be expressed by one matrix element and 10 observables.
For fixing the remaining matrix element one needs only one additional 
observable, so that indeed a complete set of 11 observables is obtained. 
An example for an explicit solution is 
\begin{eqnarray}
T_2&=&\frac{-f_{L}^{2 1} + f_{L,\,y_1}^{2 1}+
  i\,( -f_{L}^{1 1} + f_{L,\,y_1}^{1 1} ) }
{f_{L,\,xz}^{1 1} + f_{L,\,zz}^{2 1}-
  i\,( f_{L,\,xz}^{2 1} - f_{L,\,zz}^{1 1} ) }\,T_1\,,\\
T_3&=&\Big(-f_{L}^{2 1} - f_{L,\,y_1}^{2 1}+
  i\,( f_{L}^{1 1} + f_{L,\,y_1}^{1 1} ) \Big)\frac{1}{4\sqrt{6}\,T^*_1}\,,\\
T_4&=&\frac{{\sqrt{2}}\,f_{L,\,x_1}^{1 0} - f_{L,\,z_1}^{2 2}+
  i\,( f_{L,\,x_1}^{2 2} + {\sqrt{2}}\,f_{L,\,z_1}^{1 0} )}
{f_{L,\,xz}^{1 1} + f_{L,\,zz}^{2 1}-
  i\,( f_{L,\,xz}^{2 1} - f_{L,\,zz}^{1 1} ) }\,\frac{T_1}{\sqrt{2}}\,,\\
T_5&=&\frac{{\sqrt{2}}\,f_{L,\,x_1}^{1 0} + f_{L,\,z_1}^{2 2}+
  i\,( f_{L,\,x_1}^{2 2} - {\sqrt{2}}\,f_{L,\,z_1}^{1 0} )}
{-f_{L}^{2 1} - f_{L,\,y_1}^{2 1}-
  i\,( f_{L}^{1 1} + f_{L,\,y_1}^{1 1} ) }\,\frac{T_1}{\sqrt{2}}\,,\\
T_6&=&\Big(f_{L,\,xz}^{1 1} + f_{L,\,zz}^{2 1}+
  i\,( f_{L,\,xz}^{2 1} - f_{L,\,zz}^{1 1} ) \Big)\frac{1}{4\sqrt{6}\,T^*_1}\,.
\end{eqnarray}
It contains 12 observables but with two conditions
\begin{eqnarray}
(f_{L,\,zz}^{2 1})^2-(f_{L,\,xz}^{1 1})^2
-(f_{L,\,zz}^{1 1})^2+(f_{L,\,xz}^{2 1})^2&=&
(f_{L}^{2 1})^2-(f_{L,\,y_1}^{2 1})^2
-(f_{L}^{1 1})^2+(f_{L,\,y_1}^{1 1})^2\,,\\
f_{L,\,zz}^{1 1}f_{L,\,zz}^{2 1} + f_{L,\,xz}^{1 1}f_{L,\,xz}^{2 1}
&=&f_{L}^{1 1}f_{L}^{2 1} - f_{L,\,y_1}^{1 1}f_{L,\,y_1}^{2 1}\,,
\end{eqnarray}
which results in a reduction to 10 observables. 
Finally, $T_1$ can be determined from $f_L ^{00}$, the eleventh observable.

%% file: proc_conclusions.tex
\section{Conclusions and Outlook }

In these two lectures, serving as a compact survey, I have
tried to convey to the reader the vast possibilities, offered by the study of
spin degrees of freedom in electromagnetic reactions. The main conclusions are:
\begin{itemize}
\item
The density matrix formalism is an adequate, convenient and economic
framework for the treatment of spin degrees of freedom.
\item
Polarization observables provide us with much more detailed information on a
reaction than just unpolarized total and differential cross sections. 
\item
The study of polarization effects allows one to test theoretical models in
much greater detail. 
\item
In particular, polarization observables are in general much more sensitive to
small but interesting dynamical effects like, for example, parity violation.
\item
In principle, the selection and determination of a complete set of observables
provides the most detailed analysis of a reaction. 
\end{itemize}
Future experimental and theoretical efforts should be devoted to:
\begin{itemize}
\item
The development of more intense polarized beams and 
targets as well as highly efficient polarimeters.
\item
Experimental studies of various polarization observables over a large range of
energy and momentum transfers.
\item
Theoretical predictions for preferentially complete sets of polarization
observables also over a large range of energy and momentum transfers in order
to provide sensitive tests for theoretical ingredients and thus guidelines for
the planning of polarization experiments.  
\end{itemize}